\documentclass[fleqn,usenatbib]{mnras}

\usepackage{newtxtext,newtxmath}
\usepackage[T1]{fontenc}
\usepackage{ulem}

\DeclareRobustCommand{\VAN}[3]{#2}
\let\VANthebibliography\thebibliography
\def\thebibliography{\DeclareRobustCommand{\VAN}[3]{##3}\VANthebibliography}

\usepackage{graphicx}	%
\usepackage{amsmath}	%
\usepackage{subcaption}
\usepackage{pdflscape}
\usepackage{float}
\usepackage{xcolor}
\usepackage{todonotes}

\newcommand{\RN}{\text{Reissner-Nordstr\"om}}

\title[Scattering of massive particles]{Scattering of massive particles from black holes and naked singularities}

\author[A. Karakonstantakis et al.]{
Angelos Karakonstantakis,$^{1}$\thanks{E-mail: \href{mailto:karakonang@camk.edu.pl}{karakonang@camk.edu.pl}}
Włodek Kluźniak,$^{1}$
Maciek Wielgus,$^{2}$
\\
$^{1}$Nicolaus Copernicus Astronomical Center, Polish Academy of Sciences, Bartycka 18, PL-00-716 Warszawa, Poland\\
$^{2}$Instituto de Astrofísica de Andalucía-CSIC, Glorieta de la Astronomía s/n, E-18008 Granada, Spain\\
}

\date{Accepted 2026 April 08. Received 2026 March 12; in original form 2025 September 28}

\pubyear{2025}

\begin{document}
\label{firstpage}
\pagerange{\pageref{firstpage}--\pageref{lastpage}}
\maketitle

\begin{abstract}
We performed a numerical study of the dynamics of massive particles orbiting black holes and naked singularities in the \RN{} geometry. We modeled a stream of particles with a constant angular momentum and with a range of energies. We then solved the geodesic equation of motion and compared the trajectories around black holes and naked singularities by tuning the charge parameter of the metric. The setup {allows us to explore the orbital dynamics relevant for} astrophysical scenarios such as tidal disruption events{, particularly for deep encounters}. We discussed differences and similarities in the orbital dynamics and deflection angles. We found that particles reflected by a black hole follow a stream-like family of orbits within a narrow range of deflection angles, whereas in the case of naked singularities particles are scattered in all directions on the plane of motion. We explained this behavior as an interplay between the presence of a centrifugal barrier at the location of the unstable circular orbit and an absorbing event horizon in the case of a black hole or a {repulsive core} in the case of a naked singularity. These qualitative differences are expected to impact the observable signatures of tidal disruption events.
\end{abstract}


\section{Introduction}\label{s:intro}

The cosmic censorship conjecture
\citep{Penrose1969} proposes that singularities arising from continual gravitational collapse are hidden behind an event horizon, that is, inside black holes (BHs). Thus, the singularity would not be detectable by an external observer. However, direct evidence for the presence of BHs requires the detection of the event horizon, from which no particles or light can escape. Evidence for the presence of event horizons in black hole candidates is strong but inconclusive \cite[e.g.,][]{AKL02,Cardoso2019,EHT6}. Several studies
have shown that naked singularities (NkS), hypothetical compact objects in which the singularity is exposed to an external observer, can also form as the final state of gravitational collapse \cite[][]{eardleySmarr1979,christodoulou1984,oriPiran1990,joshiBook1993,giamboGiannoniMagliPiccione2004,JMN11}. Therefore, studying various aspects of NkS has been a subject of great theoretical interest, particularly in the framework of testing their existence. 

The recent remarkable advances in the angular resolution achieved by the Event Horizon Telescope (EHT) allowed for reconstructing event horizon scale images of M87$^\ast$ and Sgr~A$^\ast$ \citep{EHT_M87_p1,EHT_M87_P4,EHT_SgrA_p1,EHT_SgrA_p3}. These images carry vital information regarding spacetime geometry and allow us to directly test theories of gravity and results from simulations. The EHT tests of the spacetime metric \citep{Kocherlakota2021,EHT6} showed that \RN{} (RN) NkS predict a significantly smaller shadow size than has been observed but they did not generally rule out the possibility of the M87$^\ast$ or Sgr~A$^\ast$ spacetime being of a NkS type. \citet{ShJ19} studied two classes of NkS and their accretion disks by comparing their images and shadows to those of BHs. They found a NkS with a photon sphere that qualitatively mimicked those of a BH. On the other hand, in the absence of a photon sphere, images of NkSs differ from those of BHs and could be distinguished through future high angular resolution observations \citep[e.g.,][]{Vincent2021, Hudson2023, Johnson2024}. Additionally, the two classes of NkS considered by \citet{Shaikh19} were different from each other, allowing for distinguishing between them through their images. 
Other tests of existence of NkS have also been proposed in the literature {that rely on the  motion of material particles} \citep[e.g.,][]{Chakraborty2017,Mishra24}. 
 {Properties of standard  accretion disks have been explored for some NkS: a routine for fitting accretion disk spectra of Kerr superspinars has been developed by} \cite{Mummery2023_10.1093/mnras/stad3532} and applied to iron line fitting \citep{Mummery2024}. {Simulations of NkSs find that strong outflows often result from accretion in NkS spacetimes} \citep{KluzniakKrajewski2025,UniyalInduMizunoWK2025}.

 {
 Unlike the works cited above, relying mostly on quasi-stationary accretion features and their associated radiation, we would like to focus on transient phenomena. In this paper we explore the differences between black holes and naked singularities in the expected trajectories of unbound test particles approaching the compact object, BH or NkS. We can think of this as Rutherford's experiment for strong gravity---a small number of the scattered particles will exhibit characteristic trajectories that are signatures of the spacetime metric in which the scattering occurs. 
 Such trajectories are most likely to be observed in tidal disruption events (TDEs).
  We use the RN geometry as a model spacetime, allowing us to move from BH to NkS solutions with a change of a single parameter of the metric.
We explore the differences between the trajectories in the BH and NkS spacetimes in the hope that if they are confirmed by fluid simulations they may inform future observations of TDEs.
 }

{
\section{Tidal disruption events}\label{s:tde}
}
{We consider the trajectories of particles approaching a BH or NkS with a fixed angular momentum and various impact parameters.} We interpret our test-particle calculations in the context of actually observed astronomical events.
A star {approaching} a supermassive compact object {may} be ripped apart by tidal forces and produce a luminous transient phenomenon  \citep{Hills75,Rees88,ZTF}---a tidal disruption event. The observational properties of 56 well-studied TDE candidates have been recently reviewed in \cite{revTDE}.
{It had been predicted that}
during a TDE fragments of the stellar material would be accreted by the black hole, producing a bright flare lasting at most a few years, {while other fragments would escape the system \citep{Rees88}. This implied a wide range of expected impact parameters for fragments of the disrupted star. However, not all predictions were borne out by observations \citep{revTDE}. For instance,}  
{
for a class of these events, the so-called ``long-lived'' TDE candidates, the} {emission} {persists for decade(s)-long timescales as observed in X-rays \citep{tdeTimescales0} and optical~/~UV \citep{tdeTimescales1,tdeTimescales2}.} Likewise, a narrower stream with a limited range of impact parameters corresponding to the size of the disrupted star now seems more likely. The dynamics of stellar tidal disruption have been modeled through numerical simulations \cite[e.g.,][]{TDE89,TDE12,TDE13,Dai13,TDE15,TDE16,SadowskiTDE} as well as with simple analytical models \citep{Liu2021}.

{To relate our geodesic calculations to astrophysical scenarios, we compare the physical scales involved in a TDE. The gravitational radius is $r_g = GM/c^2 = 1.48 M/M_\odot\,$km
where $M$ is the mass of the compact object. For example, for Sgr~A$^\ast$, with $M = 4.3 \times 10^6 M_\odot$ \citep{EHT_SgrA_p1,gravity22,gravity23}, the diameter of a solar-type star ($R_\odot = 7.0 \times 10^{10}\,$cm) is a sizable fraction of the gravitational radius, $2 R_\odot / r_g = 0.22$. For a massive star of $100 M_\odot$, assuming a standard main-sequence mass-radius relation $R_\star \propto M^{0.8}$ \citep{stellar_relationships,StellarReview,StellarReference}, the stellar radius is significantly larger, $R_{\star} = 40 R_\odot$. In this case, the ratio of the stellar diameter to the Sgr~A$^\ast$ gravitational radius is $2 R_\star/r_g =8.8$. This implies that the debris stream from such a disruption would span a wide range of impact parameters in units of the gravitational radius, sampling a broad portion of the effective potential.}

{In contrast, for the supermassive black hole\footnote{Here we follow the traditional astronomical terminology, even though we mean a compact object that may be a black hole, or a naked singularity.} in M87 with $M = 6.5 \times 10^9 M_\odot$ \citep{EHT_M87_p1}, the scaling even for a massive star is significantly lower, with $2 R_\star/r_g = 0.0058$. For a sufficiently heavy central compact object the associated tidal radius may be very small in the mass units of the central object, hence TDEs can be used to probe the strong gravity regime \citep{Stone2019}.}
 
{The discovery of an off-nuclear TDE \citep{Sfaradi_2025} raises the possibility that in the future the range of masses of the compact object disrupting the observed TDEs may be extended, perhaps down to intermediate black hole masses ($M\sim10^3M_\odot$). For such objects, the range of impact parameters in a single TDE would be much higher than for supermassive compact objects. For an intermediate mass black hole ($M=10^4 M_\odot$), even a solar-type star has a diameter significantly larger than the horizon scale ($2R_\odot/r_g \approx 100$). For Sgr~A$^\ast$, a solar star covers $2R_\odot/r_g \approx 0.2$ while a $100 M_\odot$ massive star covers $\approx 10$. In contrast, for M87$^\ast$, even a massive star covers only a fraction of a percent ($\approx 0.006$) of the gravitational radius. This implies that while some streams are narrow, sampling only a small portion of the potential, others can be wide enough to sample the entire transition from capture to scattering. Our conclusions that BH streams remain narrow except near the potential peak are robust for $\Delta y_0$ up to several tens. For naked singularities, a wide range of outgoing angles is expected statistically for wide incoming streams or many such encounters.}

\section{Reissner-Nordstr\"om Metric}
\label{s:RN-metric}

\RN{} (RN) spacetime \citep{RN16,RN18} is a static, spherically symmetric solution to Einstein's field equations. It is described by two physical parameters: 
\emph{mass} \(M\), and \emph{charge} \(Q\). 
The line element of the RN spacetime,
in spherical coordinates \((t,r,\theta,\phi)\) with signature \((-, +, +, +)\), is
\begin{equation}
    ds^2=-c^2d\tau^2=-f(r)c^2dt^2+\frac{1}{f(r)}dr^2+r^2(d\theta^2+\sin^2\theta d\phi^2),
\label{eq:metric}
\end{equation}
where \(\tau\) is the proper time and \(t\) the time coordinate measured by a stationary clock at infinity. The metric function
\begin{equation}
f(r)=-g_{tt}(r)= 1- \frac{r_{\mathrm S}}{r}+ \frac{r_{Q}^2}{r^2} 
\label{eq:metricoef}
\end{equation}
involves two parameters,
the Schwarzschild radius
\(r_\mathrm{S}=2GM/c^2 = 2 r_g\) and the characteristic length scale of charge \(r_{ Q}=G^{1/2}Q/\left( c^2\sqrt{4\pi\epsilon_0}\right)\), with \(\epsilon_0\) denoting the dielectric constant. In units where \(G=c=4 \pi \epsilon_0=1\), which we use throughout the rest of the paper, the metric function can be written as 
\begin{equation}
    f(r)=1-\frac{2M}{r}+\frac{Q^2}{r^2} \, ,
\label{eq:metricoeff}
\end{equation}
and one can use the dimensionless charge $q=Q/M$.

When the 
charge $Q$ is set to zero, the metric
is reduced to that of the Schwarzschild black hole. %
Location of the event horizon can be calculated by finding the outer root of the metric function \(f(r_\mathrm{h})=0\), leading to
\begin{equation}
    r_\mathrm{h} = M + \sqrt{M^2 - Q^2}\, ,
\label{eq:horizons}
\end{equation}
when $Q<M$. In the Schwarzschild case of \(Q=0\), the event horizon is at \(r_\mathrm{h}=2M=r_\mathrm{S}\). %
The case \(M = Q\)
is known as the extreme black hole RN solution. In these cases, the singularity remains hidden by an event horizon. 

When \(Q > M\) there is no event horizon, and the resulting geometry describes a NkS. A particularly interesting property of RN NkS is that gravitational attraction vanishes on a spherical surface of radius \( r_0 = Q^2 / M\) \citep[]{Pugliese11}. In fact, gravity is repulsive inside this ``zero-gravity'' sphere,  and this  allows non-rotating fluid atmospheres to levitate around the RN NkS, and thus to cloak it with a spherical shell \citep{VK23}. Rotating fluid figures of equilibrium have the topology of a torus \citep{Mishra24}. This zero-gravity sphere is present in many other, {but not all}, models of NkSs. Hence, we expect the presented results to hold qualitatively in a wider range of geometries, although non-vacuum solutions without the presence of a repulsive core can be constructed. {For example, the JMN-1 and JMN-2 models \citep{JMN11} describe equilibrium configurations from gravitational collapse where the interior is filled with fluid, and the gravitational potential remains attractive down to the singularity. In contrast, other vacuum naked singularity solutions, such as the Janis-Newman-Winicour (JNW) metric \citep{JNW1968, Virbhadra1997}, typically exhibit a divergent repulsive potential barrier for particles with angular momentum, similar to the RN case discussed here.}

\begin{figure} 
\centering
    \begin{subfigure}{\columnwidth}
        \includegraphics[width=\textwidth]{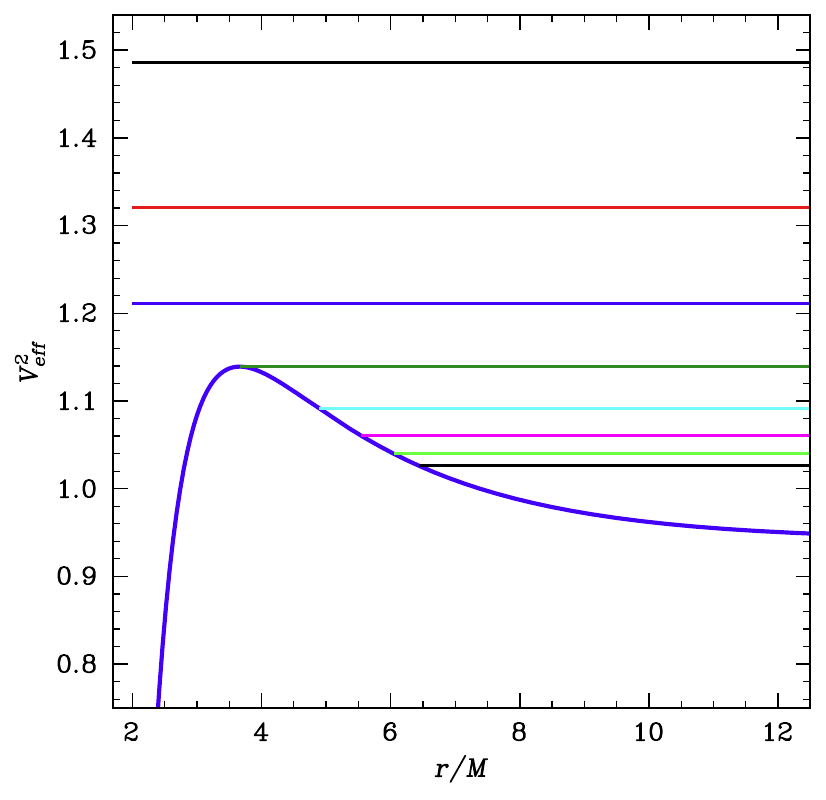}
    \end{subfigure}
\hfill
    \begin{subfigure}{\columnwidth}
        \centering
        \includegraphics[width=\textwidth]{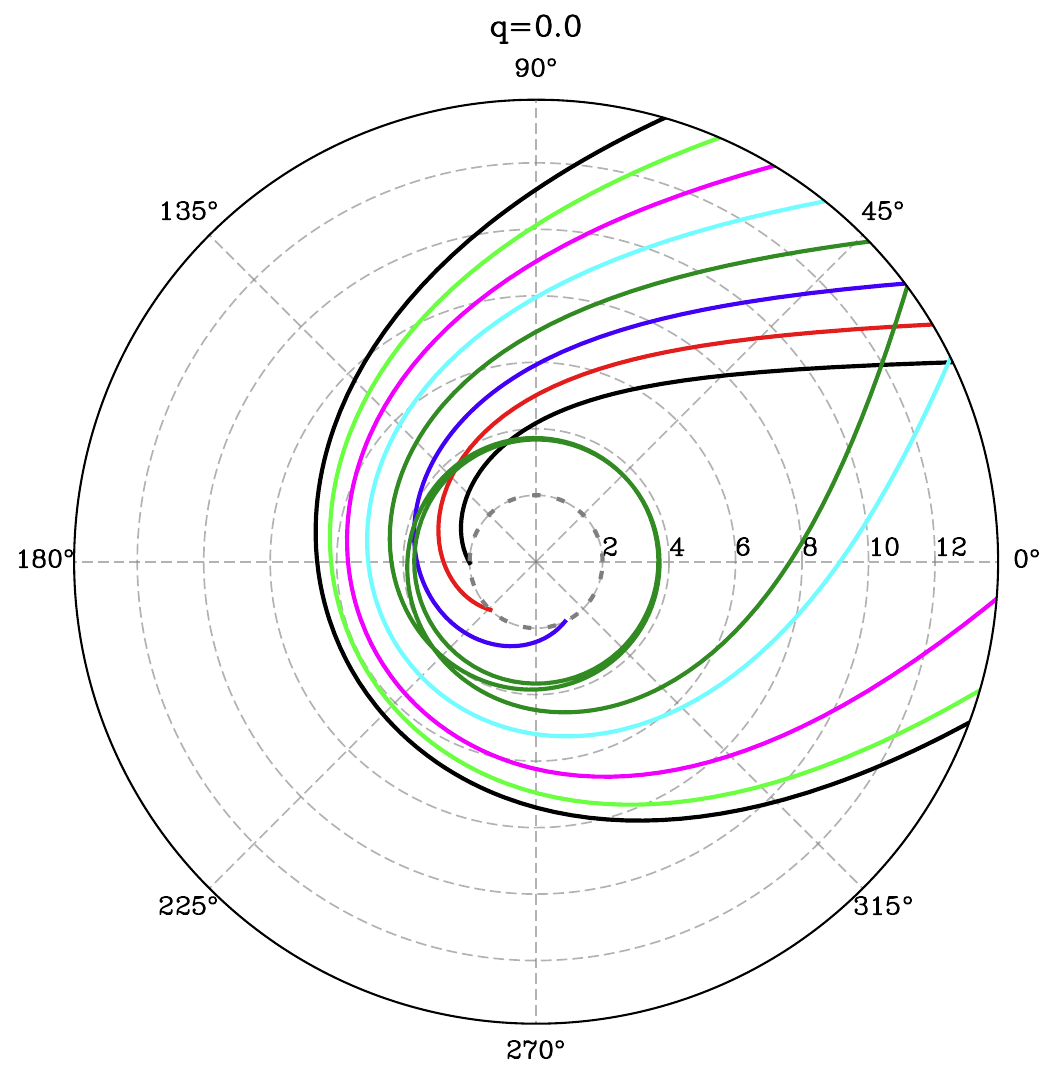}
    \end{subfigure}
\caption{The effective potential (top panel) and the corresponding time-like geodesic orbits around a Schwarzschild BH (bottom panel). All particles have specific angular momentum \(\ell=4.5M\) and their specific energy \(\varepsilon^2\) is shown with the horizontal lines in the top panel of the figure. {The horizontal lines in the top panel (from the lowest to the highest)} correspond to the trajectories in the bottom panel (solid lines of the same color, {from top to bottom, respectively, in the first quadrant)}.
}
\label{fig:constantL-Sch}
\end{figure}

\section{trajectories \& initial conditions}
\label{s:init}

We assume the RN geometry as a simple model of a spacetime and solve the geodesic equation to study the orbital dynamics of massive test particles approaching the central object. The geodesic equation reads
\begin{equation}\label{eq:geo}
\frac{d^2 x^k}{d \tau^2} + \Gamma^k_{ij} \frac{d x^i}{d \tau} \frac{d x^j}{d \tau}= 0 \ ,
\end{equation}
where $u^i={d x^i}/{d \tau}$ is the four-velocity of the particle and \(\Gamma^k_{ij}\) are the Christoffel symbols of the second kind, which can be calculated for a general metric tensor $g_{ij}$ using
\[\Gamma^k_{ij} = \frac{1}{2} g^{kl}(\partial_j g_{il} + \partial_i g_{lj} - \partial_l g_{ji}) \ .\]
The behavior around RN BHs is qualitatively similar to the Schwarzschild case, with only quantitative differences. However, as we demonstrate below, trajectories around NkSs are qualitatively different, which should have a significant impact on the course of TDEs. 

{We are interested in test particles coming in from ``infinity'', and so restrict considerations to unbound particles. In practice,}
we consider time-like geodesics initiated at a large distance from the compact object, with a Cartesian coordinate \( x_0=5000\,r_g\). The particles have a fixed specific angular momentum value \(\ell\) and an initial velocity $\varv_0$. The impact parameter, \(y_0\) is inversely proportional to $\varv_0$. In the limit of small $M/r$ we have
\begin{equation}
y_0 = \frac{\ell}{\varv_0} = \frac{\ell}{\sqrt{\varepsilon^2 -1}} \ ,
\label{eq:impact_par}
\end{equation}
from which we can evaluate initial $\varv_0$ for each impact parameter $y_0$.
The velocity components of the particles at the distant initial location are defined as:
\[u^r = -\varv_0 \cos{\phi}\, u^t \ , \] 
\[u^\phi = \frac{\varv_0 \sin{\phi}}{R_0} u^t \ ,\] 
 where \(\phi\) is the polar coordinate of the initial location $\phi = \arctan(y_0/x_0)$, and \(R_0 = \sqrt{x_0^2 + y_0^2}\) is the initial radial distance. Together with the normalization \(u_i u^i = -1\) this fully specifies the initial four-velocity.

Hence, by exploring a range of impact parameters $y_0$, we can simulate a stream of matter arriving from a distance with a constant specific angular momentum  $\ell$ and varying specific energy $\varepsilon$ as a model of TDE debris.

 In a spherical spacetime we can consider motion in the equatorial plane of $\theta = \pi/2$ without any loss of generality. The planar motion of a test particle is then described with two constants of motion, specific angular momentum \(\ell = u_\phi = g_{\phi \phi} \,d \phi/d\tau\) and specific energy \(\varepsilon =-u_t = - g_{tt}\,d{t}/d\tau\). {This allows a reduction of the problem to that of one dimensional motion in an effective potential.}

The radial equation of motion for a massive, neutral (uncharged) particle is 
{a first order differential equation that has the form of an energy conservation equation}
\citep[e.g.,][]{Chandrasekhar83}:
\begin{equation}
  (d{r}/d\tau)^2 + V_\ell^2 (r) = \varepsilon^2\ , 
\label{eq:eom}
\end{equation}%
with the effective potential
\begin{equation}
  V_\ell^2 (r) = f (r)\left[1 + \left(\frac{\ell}{r}\right)^2\right]\ ,
\label{eq:Veff}
\end{equation}
where $f$ is the metric coefficient of Eq.~\ref{eq:metricoeff}.

\begin{figure} 
    \centering
    \includegraphics[width=\columnwidth]{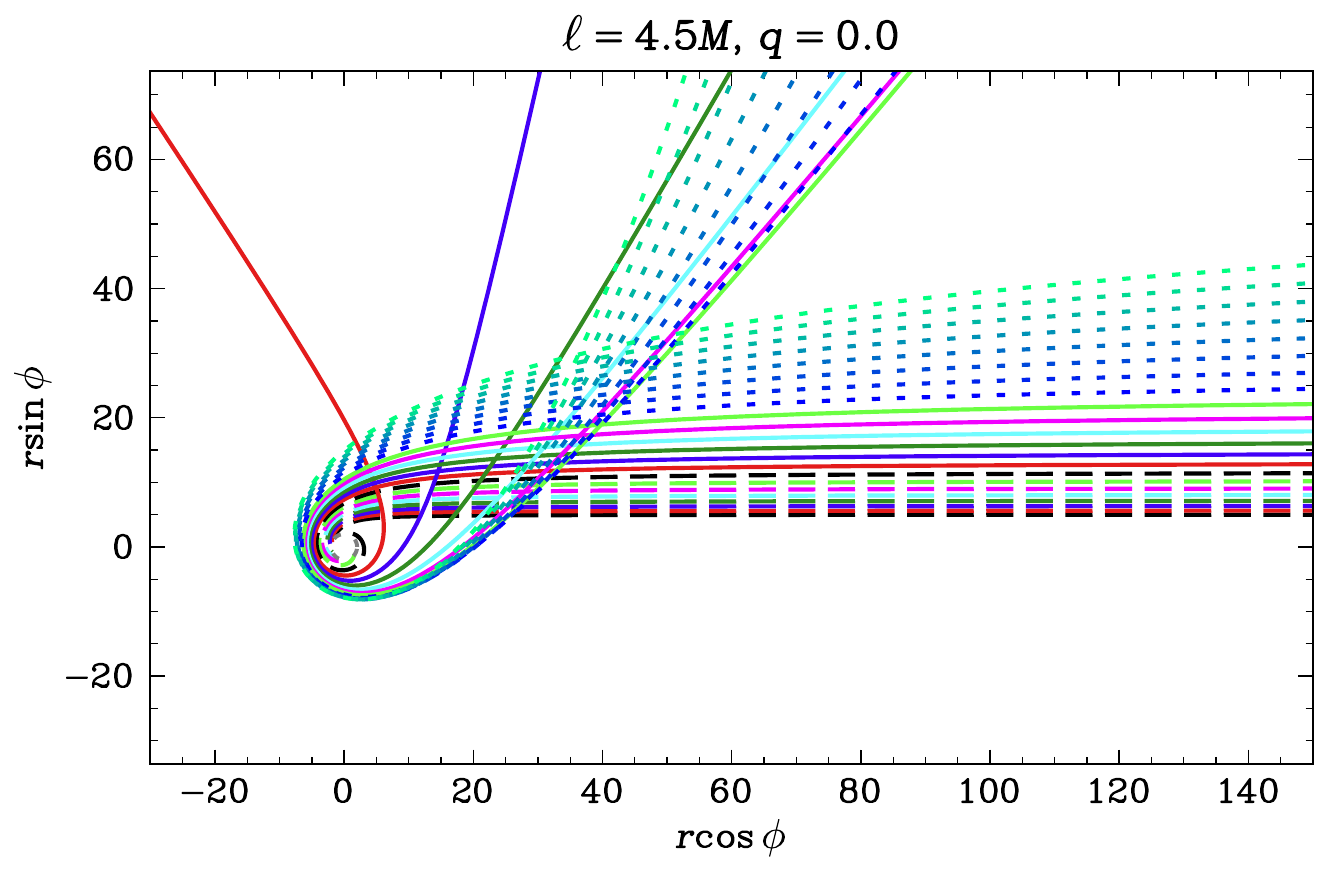}
    \caption{Trajectories for Schwarzschild BH with the same initial set up as in Fig.~\ref{fig:constantL-Sch} but with a logarithmic distribution of impact parameter between $y_0 / M < 100$. {For higher impact parameters, the orbits converge toward the $\varepsilon \to 1$ limit (see Eq.~\ref{eq:impact_par}), resulting in trajectories that become visually indistinguishable as the energy reaches its asymptotic value.} {In this figure, dashed lines correspond to high-energy (low impact parameter) orbits that cross the centrifugal barrier and are captured by the BH. Solid lines represent trajectories with energy near the critical value \(\varepsilon_{\rm crit}^2 = V_\ell^2(r_{\rm uco})\), showing large deflection angles that decrease as the impact parameter increases. Dotted lines indicate trajectories where the deflection angle trend reverses and begins to increase again, moving toward the second quadrant. This non-monotonic behavior leads to an ``excluded solid angle'' or splashback avoidance cone, as discussed more extensively in Subsection~\ref{ss:signatures}.}}
    \label{fig:constantLbunch-Sch}
\end{figure}

\section{black hole {scattering}}
\label{bhscatter}

{We begin by discussing the familiar case of timelike test-particle trajectories for a Schwarzschild BH. 
The effective potential has at most one maximum and an associated minimum at $r=r_{\rm uco}$ and $r=r_{\rm sco}>r_{\rm uco}$, respectively, corresponding to an unstable circular orbit and a stable circular orbit.}
For low values of specific angular momentum, {lower than that in the marginally bound orbit ($\ell < 4 M$ in the Schwarzschild case),} 
the effective potential {is always less than the energy of
unbound particles, $V_\ell^2 (r) <1\leq\varepsilon^2$ for all $r$, hence there are no turning points and all incoming particles will penetrate the BH event  horizon.} 
As the specific angular momentum increases,
{incoming particles with energy lower than the maximum value of the potential have a turning point at $r$ such that, $\varepsilon^2 = V_\ell^2 (r) $. The potential peak associated with the maximum} effectively acts as a centrifugal barrier that reflects infalling particles with \(\varepsilon^2 < V_\ell^2 (r_\textrm{uco}) \) back to infinity
{ while infalling test particles with \(\varepsilon^2 > V_\ell^2 (r_\textrm{uco}) \) are absorbed by the BH } \citep{GRAV-MTW}.

For a RN black hole the results will be qualitatively similar. 
Particles with energies below {the energy of the} maximum {in the potential} will be reflected, while those with higher energies will be captured by the BH. Qualitatively, the behavior of the effective potential for Schwarzschild and RN BHs is similar. As the charge-to-mass ratio \(q\) increases, the event horizon moves inward {monotonically} from \(r_{\rm h}=2M\) for \(q=0\), to \(r_{\rm h}=1M\) {for the extremal RN black hole} (\(q=1\)). {The location of the innermost stable circular orbit (ISCO), which corresponds to the marginally stable circular orbit, also decreases with increasing charge. For the Schwarzschild case ($q=0$), the ISCO is at $r_{\text{isco}} = 6M$ with specific angular momentum $\ell_{\text{isco}} = 2\sqrt{3}M \approx 3.46M$. For $q=0.5$, we find $r_{\text{isco}} \approx 5.61M$ ($\ell_{\text{isco}} \approx 3.34M$), and for $q=0.9$, the ISCO moves further inward to $r_{\text{isco}} \approx 4.51M$ ($\ell_{\text{isco}} \approx 2.99M$). In both cases, the qualitative features of the effective potential remain similar to the Schwarzschild case, but the stability threshold is shifted closer to the horizon as the electrostatic repulsion compensates some of the gravitational attraction. The location of the peak of the centrifugal barrier (\(r_\mathrm{uco}\)) also decreases accordingly.
Similarly, the location of the marginally bound unstable circular orbit, which corresponds to the peak of the effective potential with energy exactly equal to the rest mass energy ($\varepsilon = 1$, or $V_{\text{peak}}^2 = 1$), also shifts inward as the charge increases. This orbit marks the critical threshold for particles arriving from infinity with zero initial velocity: those with slightly higher impact parameter will be scattered, while those with lower impact parameter will be captured. For the Schwarzschild case ($q=0$), this marginally bound orbit is located at $r_{\text{mb}} = 4M$ with $\ell_{\text{mb}} = 4M$. As we increase the charge to $q=0.5$, the radius decreases to $r_{\text{mb}} \approx 3.74M$ ($\ell_{\text{mb}} \approx 3.87M$), and for $q=0.9$ it moves as far in as $r_{\text{mb}} \approx 2.99M$ ($\ell_{\text{mb}} \approx 3.50M$).}

{To illustrate scattering in} Schwarzschild geometry, we choose a value \(\ell=4.5M\), for which a centrifugal potential barrier exists with a maximum value of the effective potential
\(V_\ell^2 (r_\textrm{uco}) = 1.14\) at \(r_\textrm{uco} = 3.66M\). 
The geodesic orbits for particles with specific angular momentum \(\ell=4.5M\) arriving with various impact parameters are plotted in Fig.~\ref{fig:constantL-Sch}. The black, red, blue trajectories (the highest 3 horizontal lines in the top panel) cross the centrifugal barrier and fall into the BH horizon at \(r=2M\), while the other trajectories {(the bottom 5 lines in the top panel)} are reflected back. Orbits with energy $\varepsilon$ near the peak of the effective potential (unstable circular orbit at $r_{\rm uco} = 3.66M$), {may} rotate around the BH many times until they are either captured or reflected (for example the dark green line of Fig.~\ref{fig:constantL-Sch}, {with a turning point near the peak of the potential in the top panel), this explains why particles that nearly clear the peak of the BH effective potential may be scattered in any direction (depending on the actual value of $\varepsilon^2$)}.

A similar plot is shown in Fig.~\ref{fig:constantLbunch-Sch}, but with additional orbits and a different visualization style. {In this figure, high-energy trajectories (dashed lines) cross the centrifugal barrier and fall into the BH. Orbits with energy $\varepsilon$ near the peak of the effective potential are shown as solid lines; as the impact parameter increases from the critical value, the deflection angle decreases from large values. Interestingly, this trend reverses for even larger $y_0$ (dotted lines), where the deflection angle begins to increase again, moving toward the second quadrant. For a fixed value of $\ell$, this non-monotonic behavior reaches a minimum value at a transition point located approximately at the midpoint between the potential maximum and minimum.
This effect is responsible for the ``excluded solid angle'' discussed in detail in Section~\ref{ss:signatures}.}

\begin{figure}
\centering
\includegraphics[width=\columnwidth]{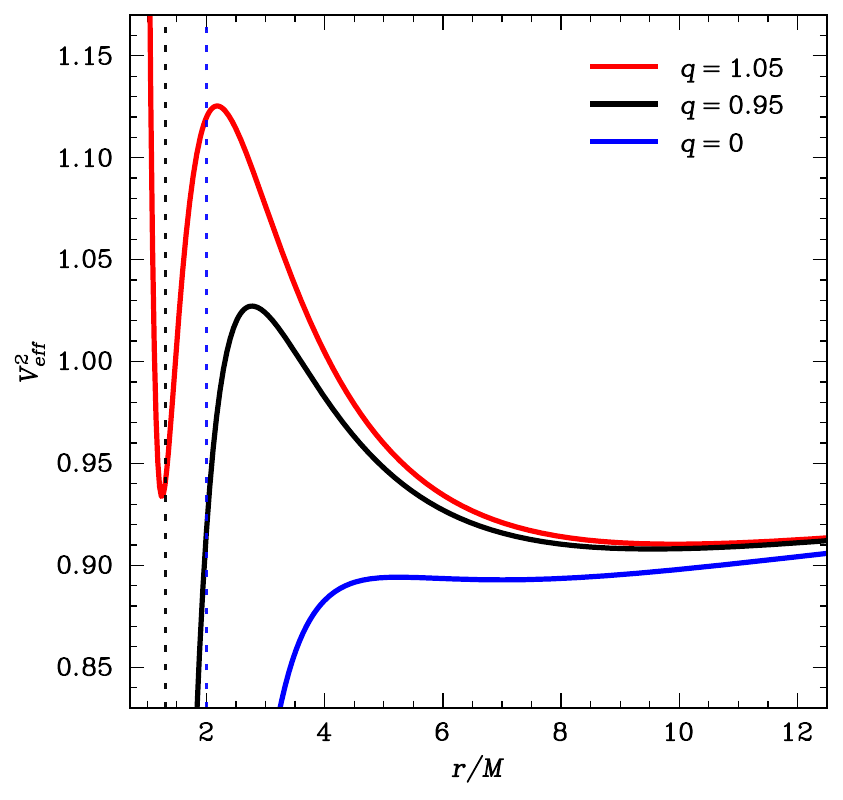}
\caption{The effective potential for time-like particles in RN metric at charge-to-mass ratio \(q\equiv Q/M=0.95\) (Charged BH; black solid curve), \(q=1.05\) (NkS; red line, topmost curve), and \(q=0\) (Schwarzschild BH; blue line, bottom most curve). Dotted vertical lines show the location of the event horizon for the two BHs. The specific angular momentum is fixed to \(\ell=3.5M\).}
\label{fig:vEff}
\end{figure}

\section{Comparison between BH and N\lowercase{k}S}\label{ss:results}
In contrast to {the attractive nature of BH gravity} everywhere outside the event horizon, 
{the gravity of an} RN NkS {includes a repulsive core.}
This fundamental difference in the spacetime structure leads to distinct orbital dynamics around BHs and NkS.  {An additional new feature is present for an RN NkS: the effective potential does not have a marginally bound orbit, nor any unstable unbound circular orbit, when $Q/M>\sqrt{32/27}\approx 1.088$ \citep{KrajewskiKluzniak2025}. This implies that for NkS with $q$ above this limit all incoming trajectories are reflected by the repulsive core. It is only for $q<1.088$ that  when approaching the unbound circular orbit trajectories may be deflected by any angle (depending on the exact value of $\varepsilon$, which must be very close to the energy in the unbound circular orbit, whether for a BH or NkS.}

Fig.~\ref{fig:vEff} shows the RN effective potential for a BH with \(Q/M=0.95\) (black line) in comparison with a NkS with \(Q/M=1.05\) (red line) and a Schwarzschild BH (blue line), {all the curves corresponding to \(\ell=3.5M\)}. For BHs the vertical dotted lines show the location of the event horizon (Eq.~\ref{eq:horizons}). In the familiar Newtonian case, the centrifugal barrier reflects all test particles with non-zero angular momentum. This is not the case for a BH, whose gravitational attraction overcomes the centrifugal barrier. For any fixed specific angular momentum of the test particle, the BH effective potential is characterized by (at most) a single maximum corresponding to an unstable circular orbit at \(r_\mathrm{uco}\), and only one minimum at $r_\mathrm{sco}>r_\mathrm{uco}$, where a stable circular orbit is located.

For \(Q = 0.95 M\) and \(\ell = 3.5 M\), as shown in Fig.~\ref{fig:vEff}, the location of the peak is at \(r_\mathrm{uco} = 2.77M\), 
with \(V_\ell^2 (r_\mathrm{uco}) = 1.027 \). Incoming test particles with energy \(\varepsilon^2 > 1.027\)
must fall into the BH, while ones with lower values of energy are ``reflected'' back to infinity at the turning point $r_\mathrm{min}$ defined by $V_\ell^2 (r_\mathrm{min})=\varepsilon^2$. For the NkS case the effective potential goes to plus infinity as $r\rightarrow0$, so there is always a repulsive core that reflects any particle that penetrates to that core, even one with zero angular momentum. However, for values $1 < Q/M < \sqrt{9/8} \approx 1.06$ the effective potential has a maximum {outside the core}, and therefore also two circular orbits (one stable and one unstable).
This is the case illustrated in Fig.~\ref{fig:vEff} for $Q = 1.05 M$ and $\ell = 3.5M$, here the peak of the effective potential at $r_\mathrm{uco}=2.18M$ with \(V_\ell^2 (2.18M) = 1.125 \) is higher than for the BH case, and is located closer to the singularity. In this case, particles with values of energy below the peak of the potential, $\varepsilon^2<1.125$, have a turning point before they reach the location of the maximum. Particles with energies $\varepsilon^2>1.125$, sufficient to ``clear'' the peak, will be reflected by the repulsive core near the zero-gravity sphere $r_0$. Thus, the main feature of scattering from NkS in geometries like the RN spacetime is that gravity, being always repulsive sufficiently close to the NkS (specifically, inside the zero-gravity sphere), reflects every sufficiently energetic test particle. This is contrary to the BH case, where, for a fixed low angular momentum, the most energetic particles are absorbed by the event horizon.

\begin{figure*}
\centering
\begin{subfigure}{\columnwidth}
\centering
\includegraphics[width=\textwidth]{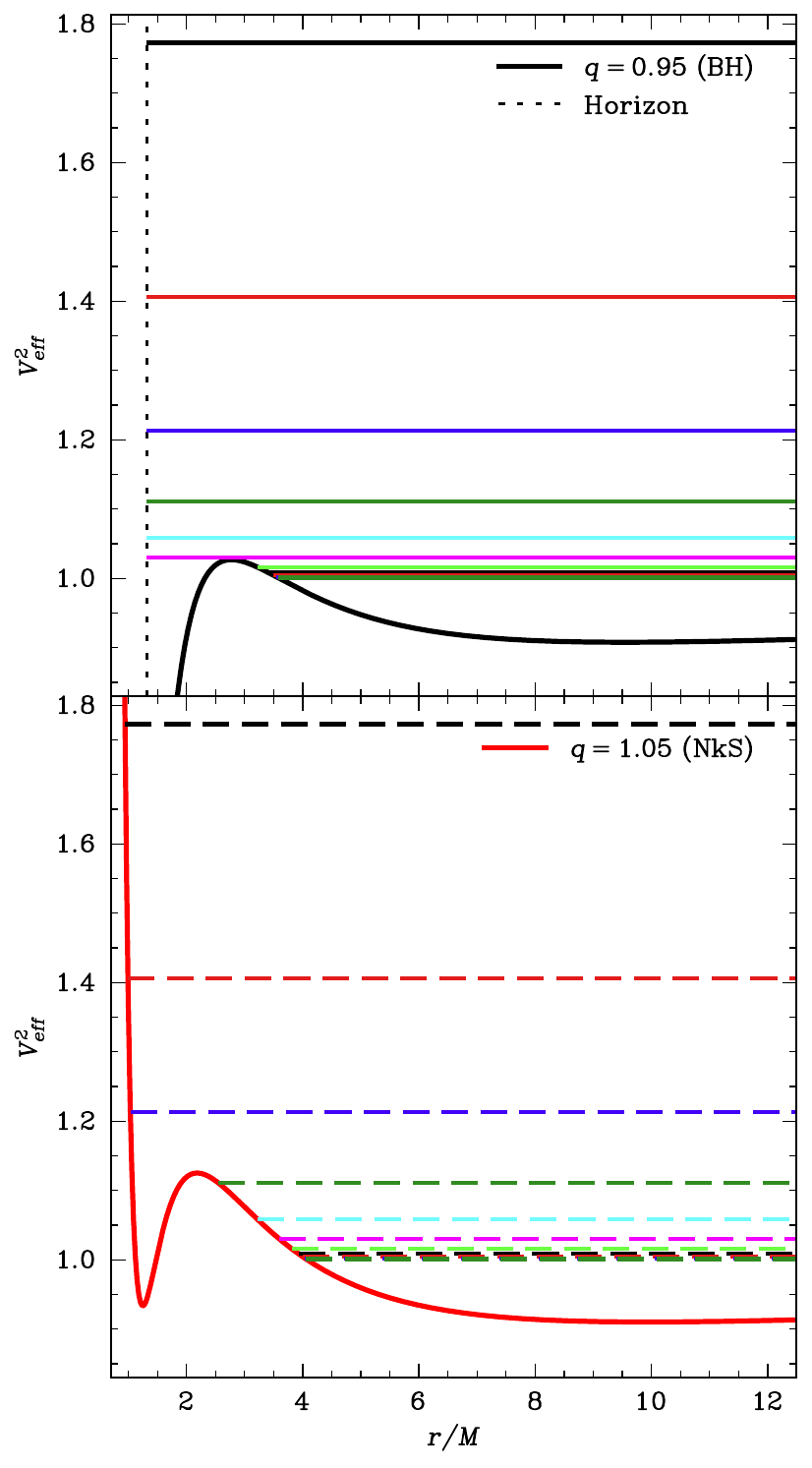}
\end{subfigure}
\hfill
\begin{subfigure}{0.9\columnwidth}
\centering
\includegraphics[width=\textwidth]{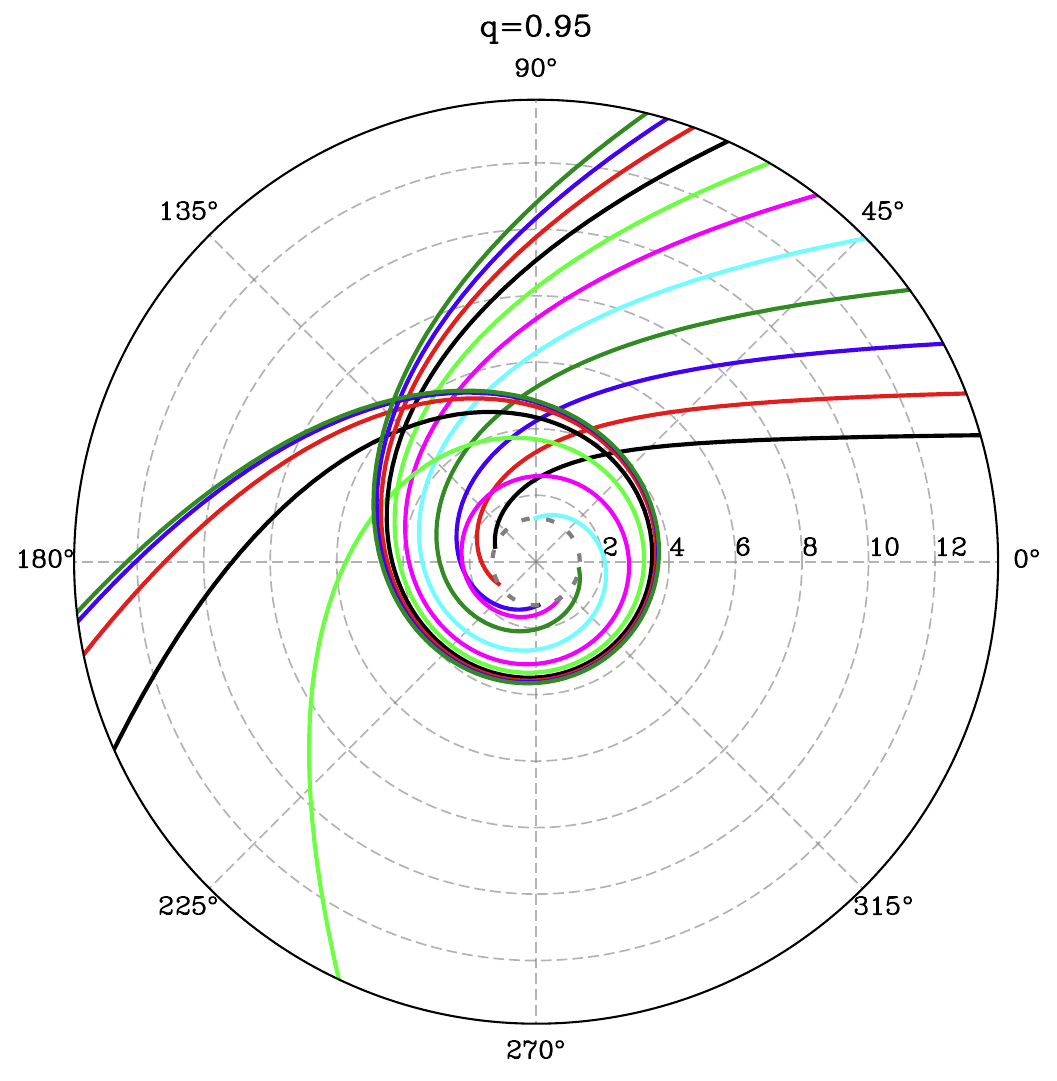}
\includegraphics[width=\textwidth]{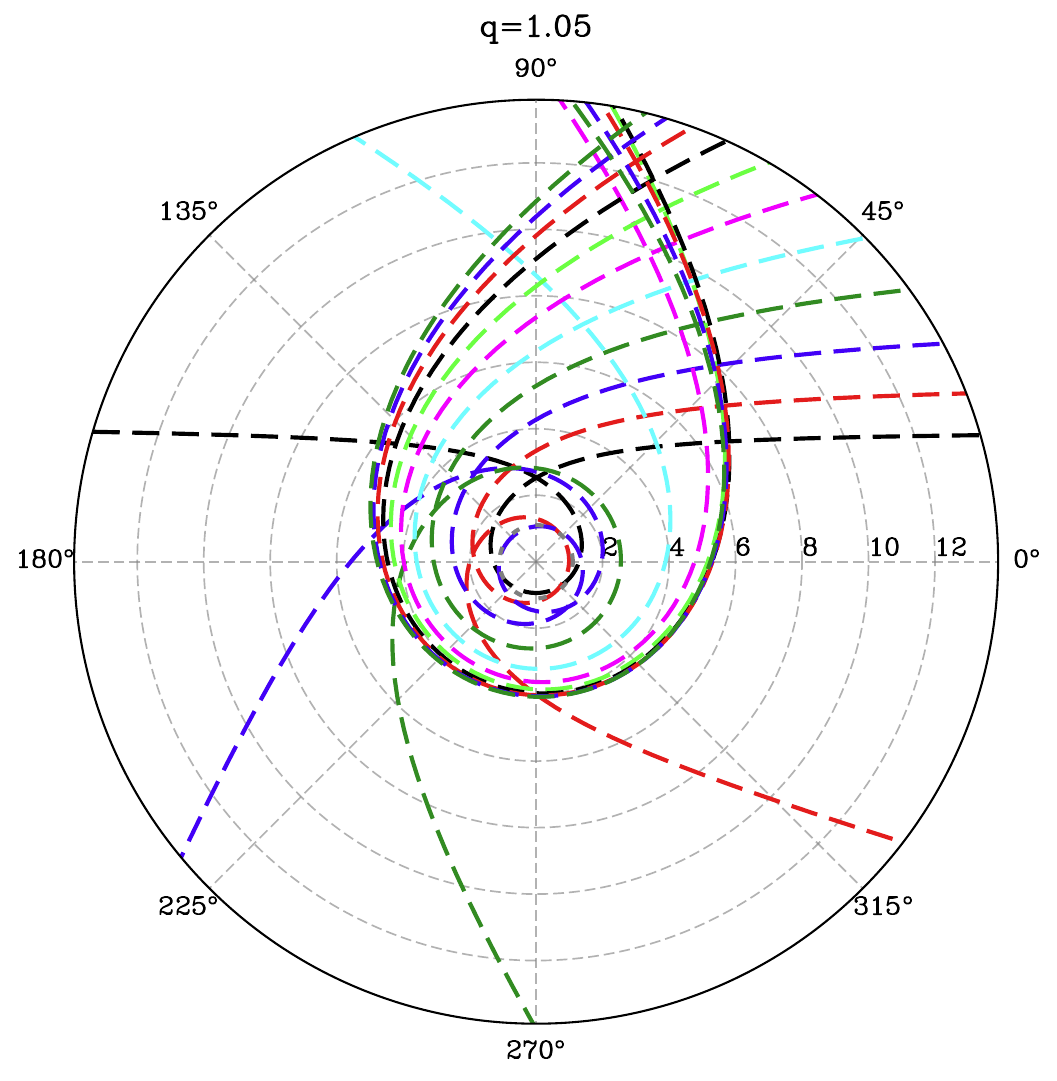}
\end{subfigure}
\caption{(\textit{left panel:}) The effective potential for particles with \(\ell=3.5M\), shown for the black hole case (\(q=0.95\), top panel) and the naked singularity case (\(q=1.05\), bottom panel). The horizontal lines correspond to the values of \(\varepsilon^2\) for the trajectories plotted in Fig.~\ref{fig:Orbits} with the same color. In the top panel, the black hole trajectories that cross the centrifugal barrier stop at the horizon radius indicated with the vertical line.
(\textit{right panel}:) Trajectories around a charged BH (\(q=0.95\), top panel) and a NkS with \(q=1.05\) (bottom panel). The colors correspond to different values of \(\varepsilon^2\) shown with horizontal lines in Fig.~\ref{fig:vEff-comp}. Specific angular momentum is fixed to $\ell = 3.5 M$.}
\label{fig:Orbits}\label{fig:vEff-comp}
\end{figure*}

\begin{figure}
\centering
\begin{subfigure}{\columnwidth}
\includegraphics[width=\textwidth]{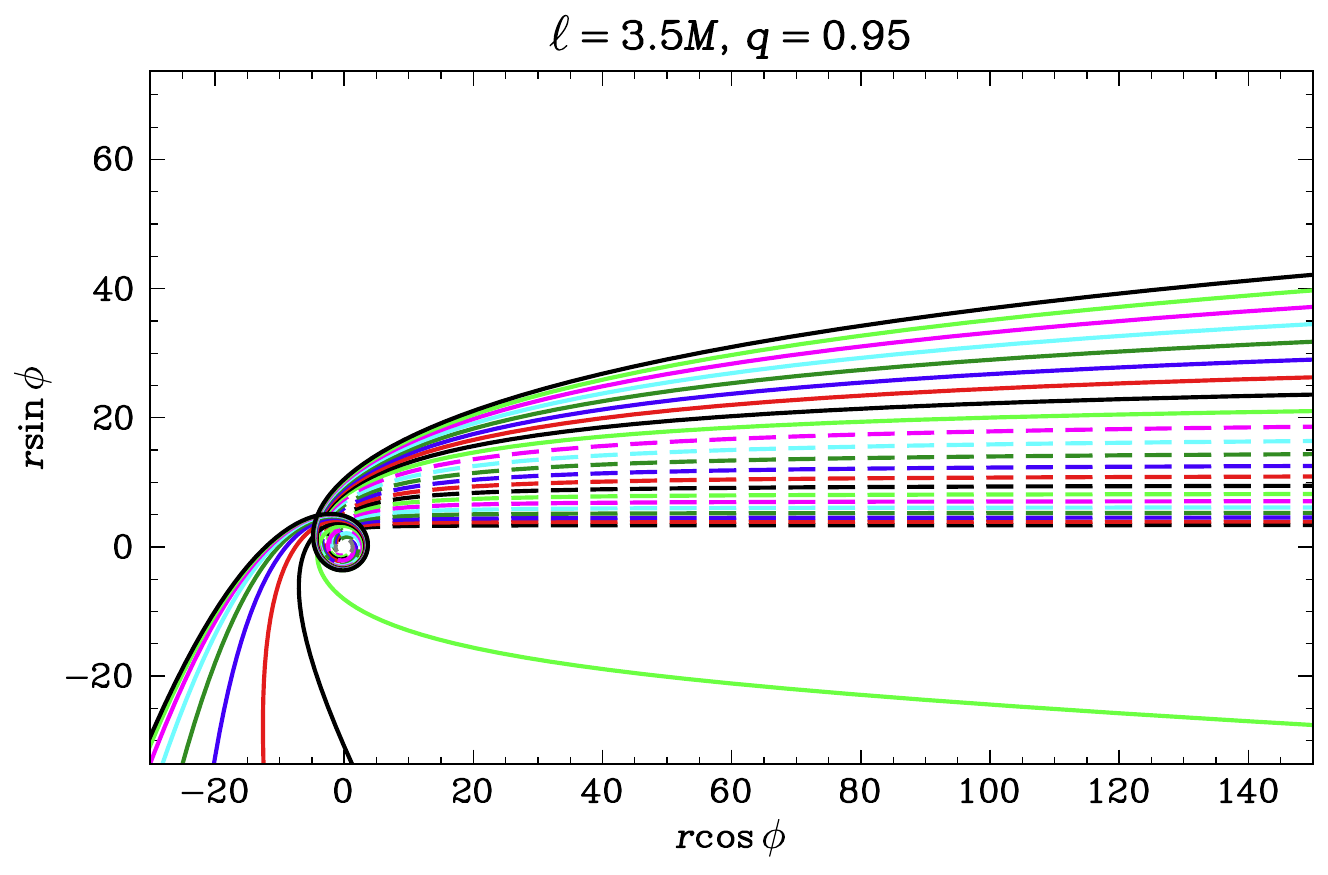}
\end{subfigure}
\hfill
\begin{subfigure}{\columnwidth}
\includegraphics[width=\textwidth]{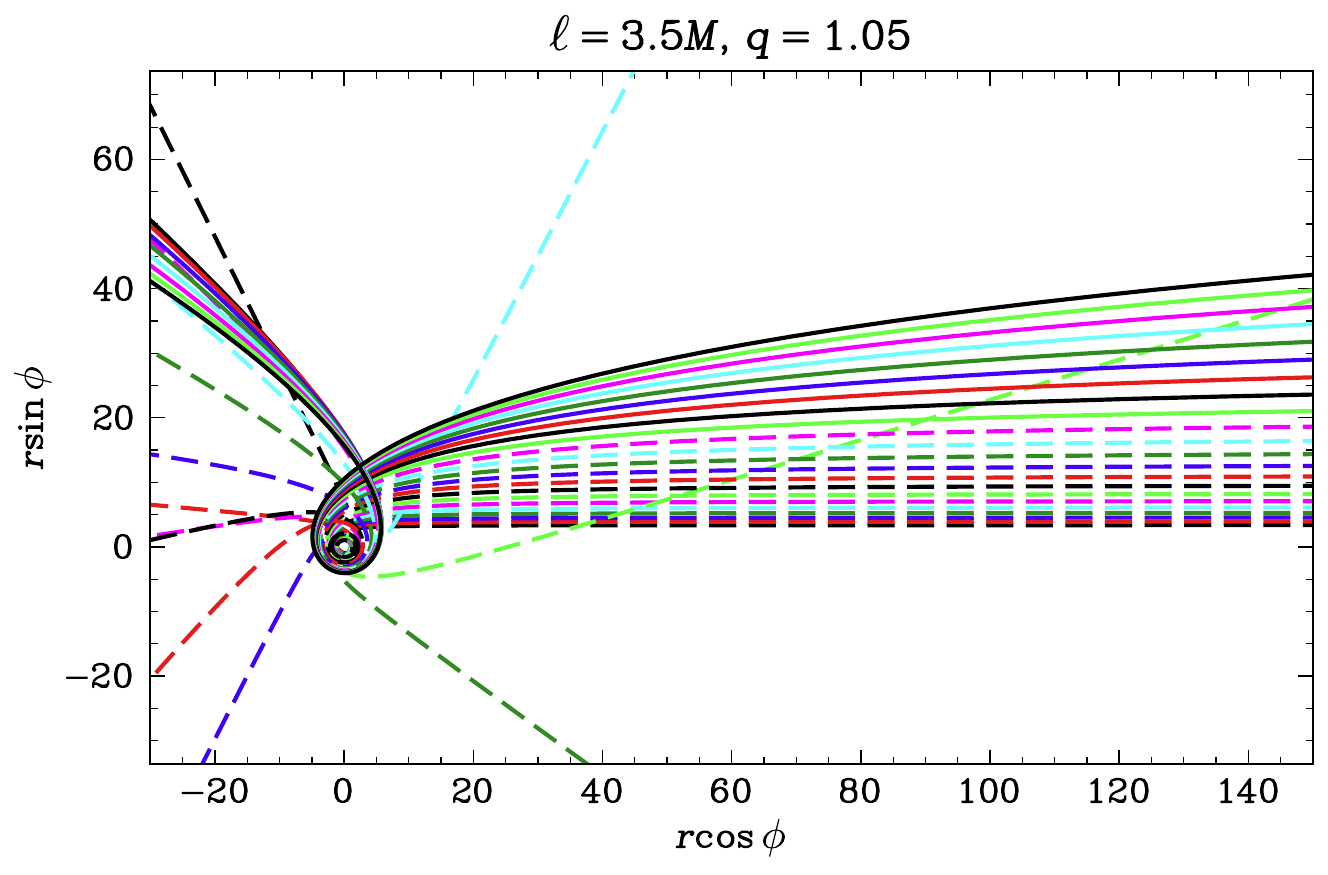}
\end{subfigure}
\caption{Trajectories around a charged BH (\(q=0.95\), top panel) and a NkS with \(q=1.05\) (bottom panel). Colors correspond to different values of \(\varepsilon^2\). Dashed lines correspond to the orbits with impact parameter \(y_0 \lessapprox 20M\) (or $\varepsilon^2 > 1.027$), that are captured by a BH, but reflected and scattered by a NkS. Specific angular momentum is fixed to $\ell = 3.5 M$.}
\label{fig:cartOrbits}
\end{figure}

We consider two values of \(q=0.95, 1.05\) as we did for the analysis of the effective potential presented earlier.  %
We set \(\ell=3.5M\) and the resulting trajectories shown in Fig.~\ref{fig:Orbits}. Before reaching the vicinity of the central singularity, the orbits exhibit similar behavior. In the BH case, orbits with a smaller impact parameter (larger $\varepsilon^2$ assuming constant $\ell$, Eq.~\ref{eq:impact_par}) are captured by the BH. As the impact parameter increases under fixed specific angular momentum, the trajectories approach the \(\varepsilon=1\) orbit, which is a consequence of Eq.~\ref{eq:impact_par}. Unlike the BH case, all orbits around the NkS are reflected back to infinity. For high impact parameters, these orbits also converge to the orbit with \(\varepsilon=1\), following Eq.~\ref{eq:impact_par}. This is also visible in Fig.~\ref{fig:cartOrbits}, where we compare these two different configurations using the same initial setup but showing additional orbits and using a finer distribution of impact parameter $y_0$ (or specific energy $\varepsilon^2$) values.

Overall, we observe a significant difference in the deflection angles between the two cases. Notably, particles reflected by NkSs are scattered in all directions. This is also evident in Fig.~\ref{fig:nearPeak}, where we show the NkS orbits with $\varepsilon^2$ values above the peak of the centrifugal barrier at \(r_\textrm{uco}\). These are the low impact parameter orbits that can be scattered at all angles. Orbits with \(\varepsilon^2\approx V_\ell^2 (r_\textrm{uco})\) will rotate around the singularity many times (e.g., the light green orbit in Fig.~\ref{fig:nearPeak}). Although the centrifugal barrier also exists in the BH case, orbits above the peak are always captured, which is the reason for this distinct feature of NkS.

{We note that the non-monotonic deflection behavior illustrated by the dotted lines in Fig.~\ref{fig:constantLbunch-Sch}, for the Schwarzschild BH, is not explicitly observed in the comparison plots for $q=0.95$ and $q=1.05$. For these specific charge-angular momentum configurations, the transition point (the minimum of the deflection angle) corresponds to energies below the $\varepsilon^2 > 1$ threshold. Since we consider only unbound trajectories in our primary analysis, the orbits at high impact parameters asymptotically converge to the $\varepsilon \to 1$ limit, where the deflection angle becomes nearly constant for a fixed $\ell$. Higher values of specific angular momenta will increase value of the potential at the peak and the same behavior in deflection angles seen for Schwarzschild BHs will be observed for Charged BHs and NkSs. Nevertheless, the effect of splashback avoidance cone still remains present, as the deflection angle for all considered unbound orbits remains ``bounded'' above a certain minimum value ($\varepsilon=1$).}

\section{{Deflection Angles}}
\label{s:scattering}

We analyze the deflection angles of the particles in this regime. {As discussed in Section~\ref{s:tde}, for a massive star disrupted by Sgr~A$^\ast$, the stream width can reach $\sim 9\,r_g$, and for an intermediate mass black hole, it can exceed $100\,r_g$. In such scenarios, the debris stream would span a wide range of impact parameters, injecting material across this entire transition region. The regime of low impact parameters (or high orbital penetration factors) corresponds to ``deep'' or ``extreme'' TDEs. In the standard black hole picture, debris in such deep encounters can be fully captured by the event horizon, suppressing the observable flare or modifying the accretion process \citep{SadowskiTDE,Stone2019}. Our results show that if the central object were a naked singularity, this ``lost'' material would instead be scattered back, potentially leading to a significantly different observational signature, such as enhanced emission or shock interactions with the trailing debris. In this section we study the deflection of a narrow stream by objects of various charge-to-mass ratios.
}

\begin{figure}
    \centering
    \includegraphics[width=\columnwidth]{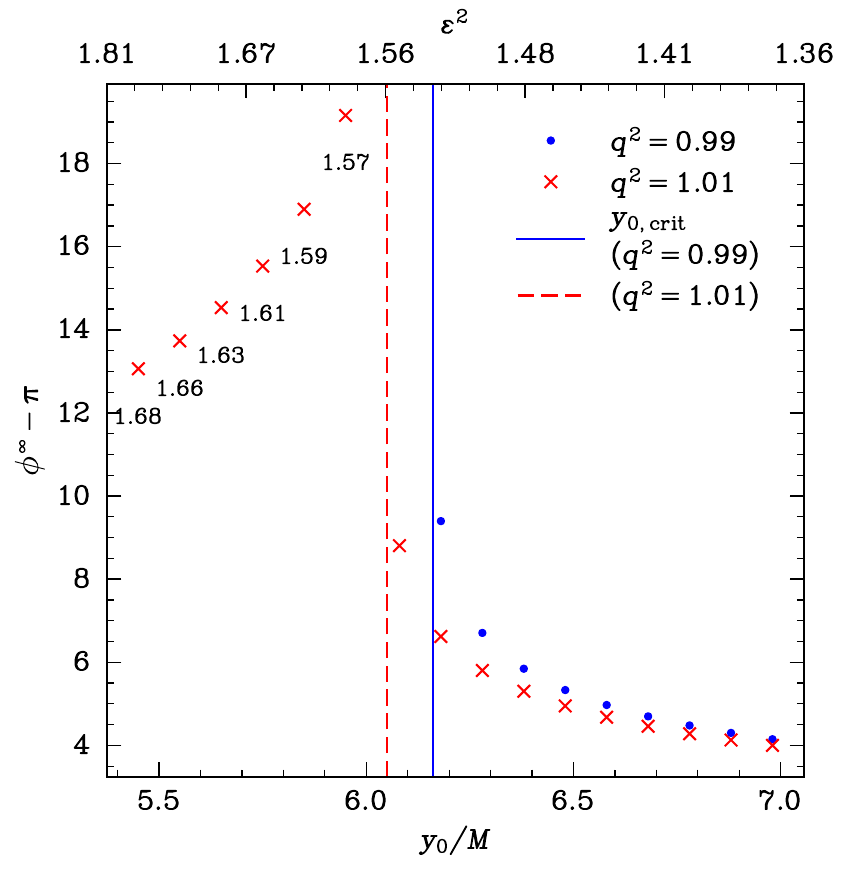}
    \caption{{Deflection angle versus impact parameter (or specific energy squared) for $q^2=0.99$ (BH, blue dots) and $q^2=1.01$ (NkS, red crosses). Vertical lines indicate the peak of the effective potential for each case. For energies below the peak (or $y_0 > y_{0,\textrm{crit}}$), the behavior is similar. Above the peak ($y_0 < y_{0,\textrm{crit}}$), BH orbits are captured (not shown), while NkS orbits are scattered. For those NkS orbits that are reflected by the repulsive core, values below the red points indicate the corresponding $\varepsilon^2$ value. The trajectories corresponding to each point are shown in the next Fig.~\ref{fig:angles}.}}
    \label{fig:phi_combi}
\end{figure}

{
First, we focused on the transition between black hole and naked singularity spacetimes by considering charge squared values of $q^2 = 0.99$ (BH) and $q^2 = 1.01$ (NkS) for a specific angular momentum $\ell = 4.5 M$. Fig.~\ref{fig:phi_combi} shows the deflection angle, calculated as $\phi^\infty - \pi$, as a function of the squared specific energy $\varepsilon^2$ or the initial impact parameter $y_0$. In this range of charge $q$, for the naked singularity, the effective potential exhibits two circular orbits (one stable and one unstable). This contrasts with the black hole case, which possesses only the unstable orbit (in the inner region). The presence of the inner repulsive core in the naked singularity case allows for stronger scattering of particles that penetrate the centrifugal barrier. 
The transition region is characterized by a narrow range of impact parameters. The vertical lines in Fig.~\ref{fig:phi_combi} indicate the critical impact parameter. The range of impact parameters where the black hole captures particles while the naked singularity scatters them spans over $\Delta y_0 = 1.5 M$. Approximately $0.5M$ of that range between $y_0 = [5.5 - \sim{6})M$ corresponds to values above (below) the critical specific energy (or impact parameter); thus, particles in this regime are absorbed by the BH but reflected by the NkS repulsive core, while the rest $\Delta y_0 \approx 1 M$ is reflected for both objects in a similar manner.
}
\begin{figure}
    \centering
    \includegraphics[width=\columnwidth]{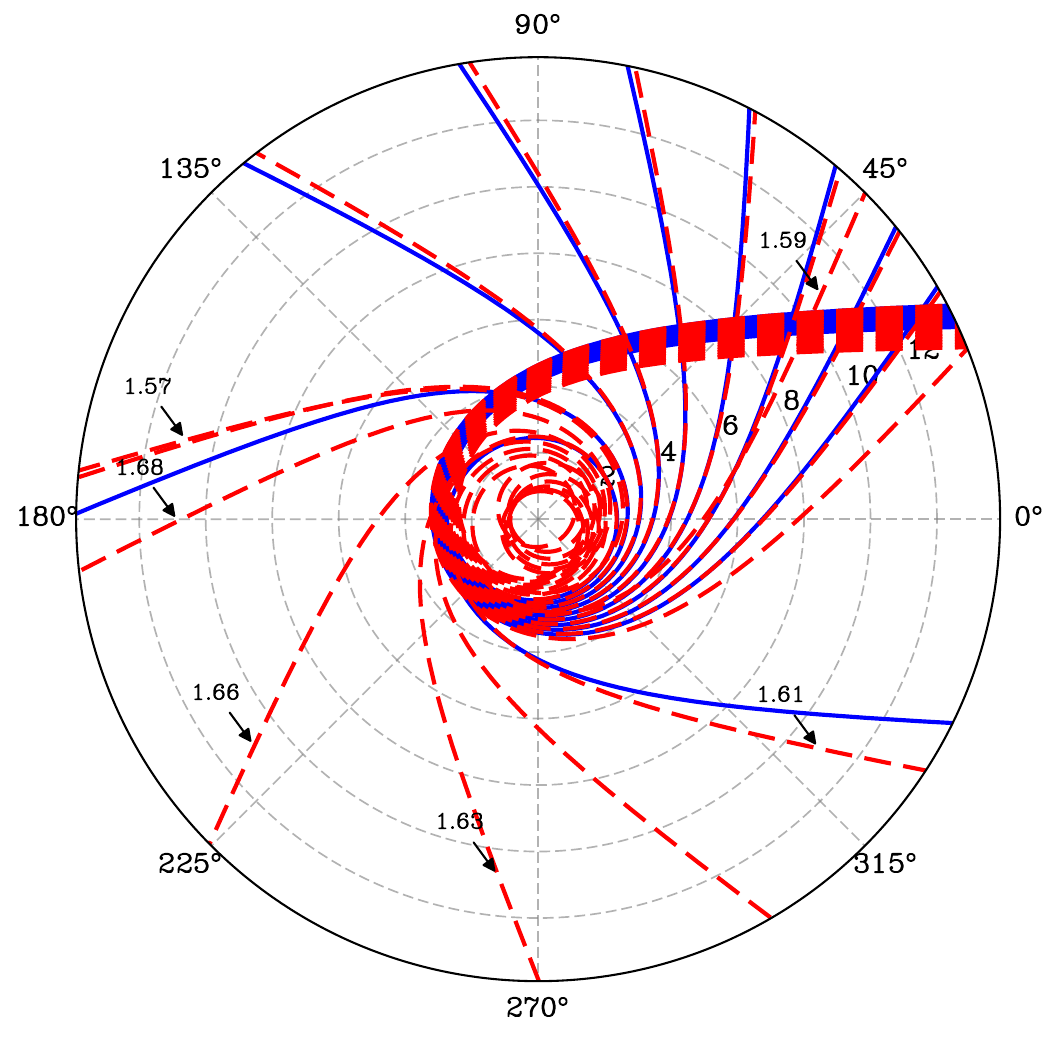}
    \caption{{Trajectories of particles in the transition region. Blue lines represent orbits reflected by the centrifugal barrier of the nearly extremal BH. Red lines show orbits reflected by the centrifugal barrier and repulsive core of the NkS, resulting in large deflection angles. Arrow annotations indicate orbits with $\varepsilon^2 > \varepsilon_{\rm crit}$ values of Fig~\ref{fig:phi_combi}.}}
    \label{fig:angles}
\end{figure}

{
The divergence in the deflection angle, which occurs as the particle's energy approaches the peak of the effective potential (corresponding to the unstable circular orbit), is further illustrated in Fig.~\ref{fig:angles}, which displays the trajectories of the particles in this transition region. While particles with energies below the peak follow similar scattering paths in both spacetimes, those with energies exceeding the peak are captured by the black hole but are scattered by the naked singularity. The labels/arrows in Figs.~\ref{fig:phi_combi} \& \ref{fig:angles} indicate the corresponding squared specific energy ($\varepsilon^2$) for those trajectories.
}

{
For energy values below the peak of the effective potential (indicated by vertical lines in Fig.~\ref{fig:phi_combi}), the orbits in both spacetimes are reflected by the centrifugal barrier and show similar deflection angles. As the energy approaches the critical value corresponding to the unstable circular orbit at the peak of the potential, the deflection angle diverges, corresponding to the particles whirling around the central object multiple times.
Crucially, for energies exceeding this critical value ($\varepsilon > \varepsilon_{\rm crit}$), the behaviors diverge. In the black hole case ($q^2=0.99$), these particles cross the barrier and are captured by the event horizon. However, in the naked singularity case ($q^2=1.01$), the particles pass through the centrifugal barrier but are subsequently reflected by the repulsive core near the singularity. These particles return to infinity with large deflection angles, effectively populating a region of the parameter space that is empty (due to capture) in the black hole scenario. Extending the black hole stream for larger impact parameters, we go away from the critical value (i.e. $y_0 \gg y_{0,\text{crit}}$). In this regime, which corresponds to the ``narrow stream'' comparison discussed in the following subsection, the black hole and naked singularity produce similar reflection signatures.
}

The range of impact parameters used in our calculation {(so far) corresponds to the region around the peak of the effective potential where this divergence occurs. This behavior is specific to the charge range $1 < q \lesssim 1.06$ where the effective potential retains the centrifugal barrier and the associated inner stable orbit. For larger charges (e.g., $q \ge 1.1$ discussed later), the barrier may disappear for the considered angular momentum, altering the scattering dynamics.}

{
If the central object is a naked singularity, the high-energy material (low impact parameter) would be scattered back into the surroundings, potentially interacting with the outer parts of the stream, whereas for a black hole, it would be lost from the system.
}

{
We also examine the scattering of a spatially confined stream with a width of $\Delta y_0 = 0.5 M$, relevant for tidal disruption events where the debris stream is narrow compared to the gravitational radius. Fig.~\ref{fig:narrow_streamNkS} compares the deflection angles for black holes and a naked singularity. For the black hole cases ($q < 1$), the stream is either captured (for low impact parameters) or reflected by the centrifugal barrier. For highly charged black holes (e.g., $q=0.95$), the reflected part of the stream is focused into a narrow range of deflection angles. Interestingly, the naked singularity case ($q=1.05$) shows a similar focusing effect for this specific range of impact parameters, scattering the stream into a confined angular sector, distinct from the broad scattering observed for wider streams. This focusing suggests that for distant encounters, the repulsive core does not necessarily lead to broad scattering, but rather preserves the stream-like morphology of the debris, provided the impact parameters remain significantly above the critical value. In an astrophysical TDE, this would imply that only the deepest-penetrating part of the stream would be broadly scattered.  %
}

\subsection{Dynamical Signatures and Observational Consequences}\label{ss:signatures}
{
Building upon the previous analysis, we now synthesize the results to outline a framework for dynamically distinguishing between black holes and naked singularities based on the observed characteristics of scattered TDE streams. The key question is: given the orbital properties of a debris stream (e.g., its energy and angular momentum) and the mass (ratio) of the central (and disrupted) object, what dynamical outcomes point exclusively to a naked singularity or could set quantitative limits to a BH charge-to-mass ratio?
Our study involved two primary discriminators: the binary outcome of capture versus reflection, and the width of the resulting scattered stream.
}

\paragraph*{Capture vs. Total Reflection:}
{
The most robust and fundamental difference lies in the fate of deeply plunging orbits. For a given angular momentum, a black hole always possesses a critical energy threshold above which all particles are captured by its event horizon. If a tidal stream has parameters that would lead to capture in a black hole spacetime (e.g., low angular momentum, $\ell < 2\sqrt3 M$ for Schwarzschild), the observation of any reflected material strongly indicates the absence of a horizon and the presence of a repulsive core, a ``hallmark'' of a naked singularity. As shown in the Appendix (Fig.~\ref{fig:appendix_combined}), even when the centrifugal barrier is absent for a naked singularity (e.g., for $\ell=2.7M$), deep encounters are still entirely reflected, %
whereas they would be entirely captured by a comparable black hole.
}

\paragraph*{Scattering Angle Distribution:}
{
The distribution of deflection angles of a stream, see details in Section~\ref{s:scattering}, provides another, more refined test.
\subparagraph*{Wide-Angle Scattering:} A key signature of a naked singularity (in the regime $1 < q \lesssim 1.06$ or for wide streams) is the potential for extremely wide-angle, almost isotropic scattering. This occurs when narrow stream particles have impact parameters near the peak of the effective potential (unstable circular orbit), leading to chaotic "whirling" orbits before expulsion. For a wide stream, even if particles interact only with the inner repulsive core, they can be scattered widely (see Fig.~\ref{fig:nearPeak}). A black hole, by contrast, would capture (part of) these same particles, leading to an absence of returned material. For a stream with an impact parameter near the critical value, we studied near-extremal BHs and NkSs (see Fig.~\ref{fig:angles}) and showed that while, both objects can cause all-angle scattering, a NkS produces stronger deflection of the material from a repulsive core.
As we recognize that this test has limitations in robustness and requires careful fine-tuning of parameters due to its sensitivity to specific conditions, we still acknowledge it as a physically distinct phenomenon. Additionally, given the specific angular momentum and impact parameter of a \textit{narrow} infalling stream, we could set qualitative constraints on the charge-to-mass ratio as we can exclude $q>1.06$ that lack circular orbits. Additional constraints could be set for charged BHs: for example if $\ell=3.5M$ the stream will be fully absorbed by the Schwarzschild BH (see Fig.~\ref{fig:vEff}) and could be (widely) scattered by a charged BH/NkS. We can constrain the charge-to-mass ratio into a definite value that produces a peak near the given stream's impact parameter range} \citep[see, e.g.,][for analytical treatments of circular orbits]{Pugliese11}.

\begin{figure} 
\centering
    \begin{subfigure}{\columnwidth}
        \includegraphics[width=\textwidth]{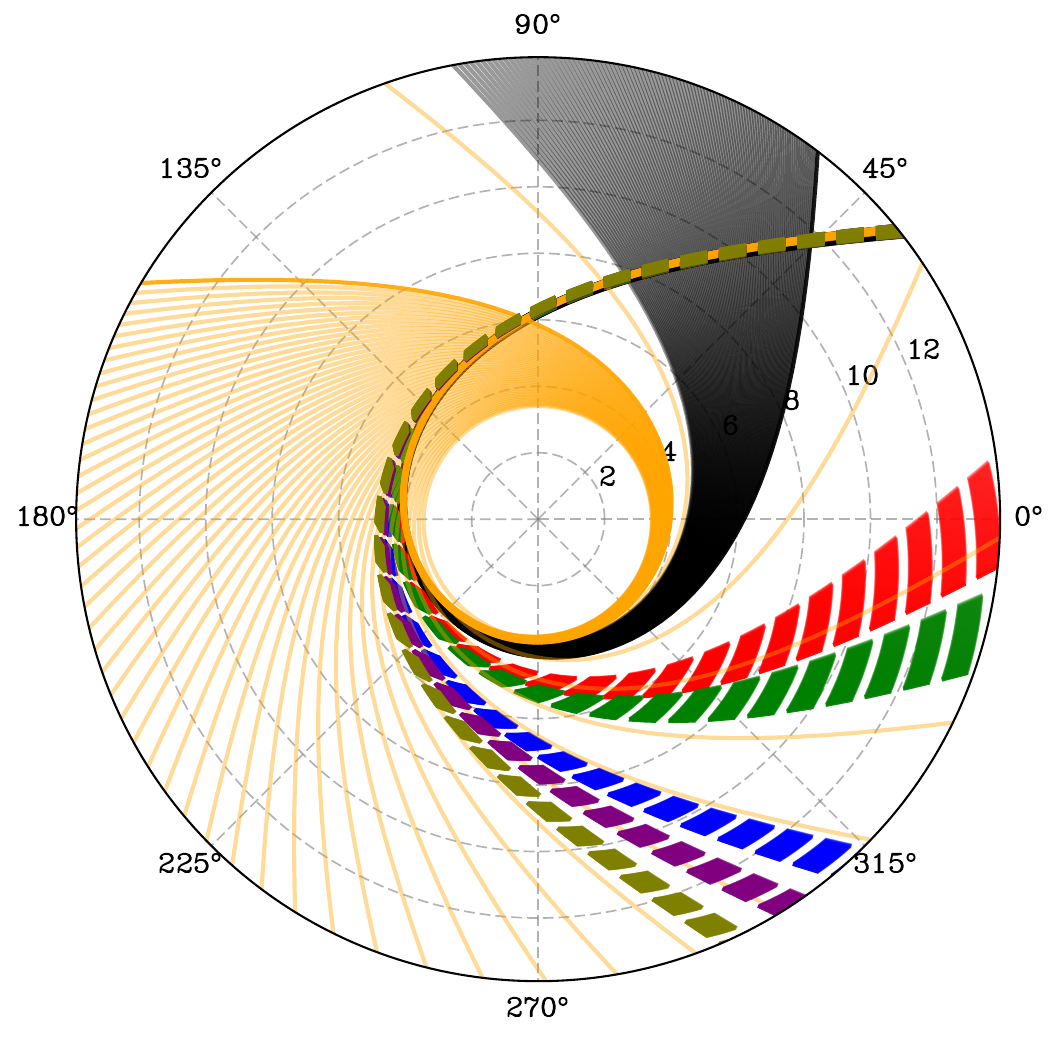}
    \end{subfigure}
\hfill
    \begin{subfigure}{\columnwidth}
        \centering
        \includegraphics[width=\textwidth]{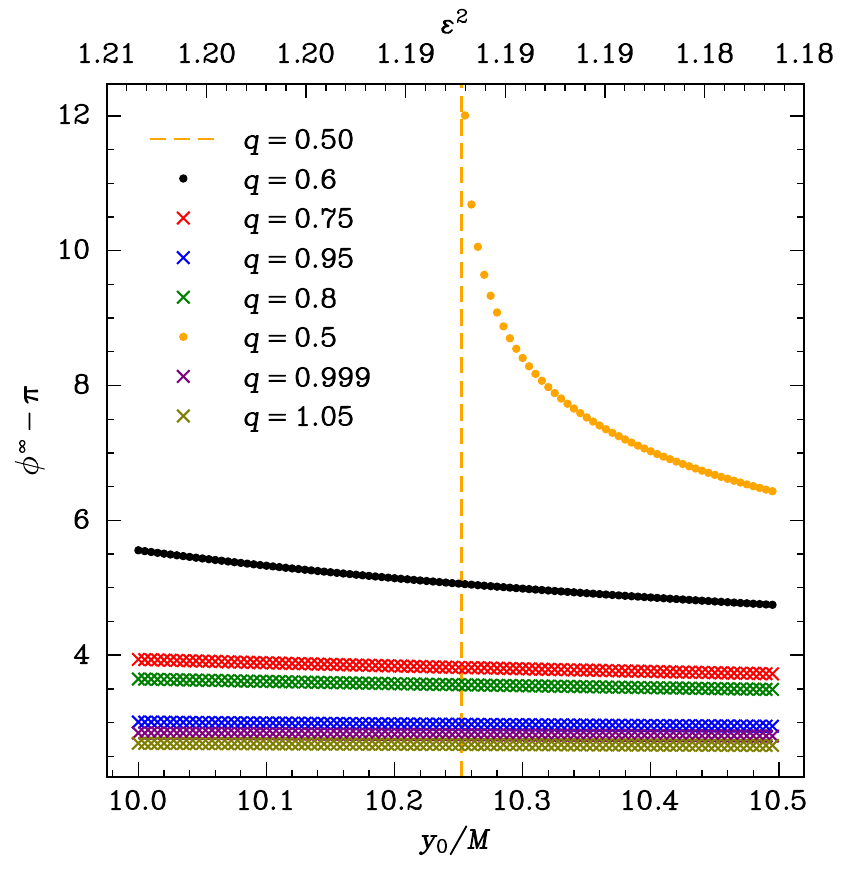}
    \end{subfigure}
\caption{{Comparison of deflection angles for a very narrow stream ($\Delta y_0 = 0.5 M$, centered at $y_0 > 10$) interacting with black holes and a naked singularity. The impact parameters correspond to orbits with $\varepsilon$ below the peak of the effective potential for all values of charges considered here apart from $q=0.5$, for which the stream is partially absorbed and broadly reflected. For the other highly charged black hole cases, we observe a focused reflection with the width of the reflected stream decreasing as the charge approaches the NkS regime. The naked singularity ($q=1.05$) also scatters the stream to a narrow span of angles, similarly to the highly charged BHs.}}
\label{fig:narrow_streamNkS}
\end{figure}

\subparagraph*{Narrow-Stream Focusing:} {For streams that do not interact closely with the potential peak (i.e., having energies well below it), both (un)charged black holes and naked singularities can reflect the debris into a similarly narrow cone (Fig.~\ref{fig:narrow_streamNkS}). In this regime, distinguishing between the two object types based solely on the scattering angle is challenging.  %
}

{Interestingly, as one moves from just below the potential peak to lower energies (larger impact parameters), the deflection initially decreases, reaches a minimum, and then begins to grow again at even larger $y_0$. This behavior implies an ``excluded solid angle'' for the Schwarzschild case (see Fig.~\ref{fig:constantLbunch-Sch}); for a given initial angular momentum, there is a cone around the point of origin that the debris never enters, as the deflection angle is always above a certain minimum value. This ``splashback'' avoidance cone exists for other values of $q$ as well, though the range of excluded angles depends on the specific geometry.}

\section{Discussion}\label{s:discussion}

In this study, we investigated the orbital dynamics of particles around black holes and naked singularities in the \RN{} spacetime. We analyzed the effective potential for massive particles with different specific angular momenta and examined the orbits around a Schwarzschild BH $(q=0)$, charged BH $(q=0.95)$, and NkS $(q=1.05)$. We focused our primary analysis on specific angular momentum and charge values for which a centrifugal barrier exists, allowing for a direct comparison between black hole capture and naked singularity scattering. 

{In the accepted picture of TDEs, post-disruption stellar debris acquire a spread of specific energy and angular momentum. Roughly half of the stellar debris is unbound and escapes the system, and the other half remains bound to the central object. In this work, we considered only unbound orbits moving toward the central singularity. For a direct and easier comparison of the trajectories with the effective potential that determines the radial motion, we set the specific angular momentum at the same value for all particles. Nevertheless, we have shown that the fundamental differences between the black hole and naked singularity cases arise from the interplay between the centrifugal barrier and the repulsive zero-gravity sphere, rather than from the specific orbital characteristics of the particles.}

{The differences in trajectories identified here could have profound implications for the interpretation of Tidal Disruption Events. In the standard BH scenario, the most energetic parts of the stellar debris are lost, but in a NkS spacetime, this material is returned to the environment with potentially very high deflection angles. This could lead to enhanced self-intersection of the debris stream and faster circularization, or perhaps more chaotic accretion signatures. We emphasize that while individual narrow streams might appear focused even around a NkS, the wide range of possible outgoing angles is true statistically when considering many such encounters or a sufficiently wide incoming stream. In other words, even if we miss the potential peak most of the time, the aggregate of many narrow scattered streams will be distributed over $2\pi$ for a NkS, whereas a BH would simply absorb the corresponding material.}

{Our analysis of the deflection angles further suggests a hierarchy of signatures depending on the charge parameter. For specific impact parameter ranges, a naked singularity can focus the scattered stream into a narrow angular sector, mimicking black hole behavior for trajectories that miss the centrifugal barrier. However, we identified a critical charge threshold ($q \gtrsim 1.088 = \sqrt{32/27}$) above which the centrifugal barrier vanishes for all unbound trajectories, leading to a scattering regime entirely dominated by the repulsive core. In this high-charge limit, the absence of wide-angle scatter despite many encounters would point toward a NkS. Conversely, if some incoming streams systematically fail to emerge, it indicates an absorbing event horizon (a BH) capturing material above the potential peak. In this context, ``black hole'' refers to a configuration with attractive gravity all the way down, potentially encompassing gravastars if the return time for material is prohibitively long or if they lack efficient splashback mechanisms.}

{We also note that the deflection angle exhibits non-monotonic behavior as a function of the impact parameter. Moving from the potential peak towards lower energies, the scattering angle first decreases to a minimum value before increasing once more. This implies the existence of an ``excluded solid angle'' or a ``splashback avoidance cone'' centered on the axis of incidence. Within this cone, no scattered material is expected for a given angular momentum, a feature present in both Schwarzschild and RN geometries with varying angular extents.}

The behavior around RN BHs is qualitatively similar to the Schwarzschild case, with only quantitative differences. As the charge-to-mass ratio increases, the radius of the ISCO, the location of the peak of the centrifugal barrier, and the event horizon are shifted inwards.

{We also explored the transition regime ($q^2 \approx 1$) and found that for ``extreme'' TDEs with deep impact parameters, the outcome is fundamentally different: while a BH captures the most energetic part of the stream, a NkS scatters it back with high deflection angles. This scattered material could interact with the trailing debris, potentially leading to inefficient circularization or distinct shock signatures \citep[e.g.,][]{Liu2021}.}

However, predicting the aftermath of the interaction of the relativistic streams is a complex problem that will likely require numerical general relativistic magnetohydrodynamic simulations of a TDE around a NkS.
{In a follow-up work, we are currently performing General Relativistic Smoothed Particle Hydrodynamics (GR-SPH) simulations of TDEs in Reissner-Nordstr\"om geometry using the PHANTOM code \citep{Price2018, Liptai2019, KarakonstantakisInPrep}. The geodesic solver developed in this study has served as a crucial tool for validating the metric implementation in our hydrodynamic code. These future simulations will allow us to self-consistently model the fluid dynamics, stream self-intersection, and accretion disk formation, building upon the dynamical insights presented here.}

\section{Conclusions}\label{s:conclusions}

We have analyzed the geodesic motion of massive particles in the Reissner-Nordstr\"om spacetime, comparing the cases of black holes and naked singularities. Our primary findings are summarized below:

{
\begin{enumerate}
    \item For BHs, particles with energy exceeding the potential barrier peak ($\varepsilon^2 > V_\ell^2 (r_\textrm{uco})$) are captured by the event horizon. Reflected particles exhibit a limited range of deflection angles.
    \item For NkSs, all particles are reflected. Those with sufficient energy to cross the centrifugal barrier are scattered by the inner repulsive core. Unlike the BH case, this leads to scattering at all possible angles due to the particles whirling around the potential peak before expulsion.
    \item Analysis of narrow streams reveals that for $q=1.05$, the naked singularity focuses the scattered debris into a confined angular sector, mimicking the behavior of highly charged black holes, provided the stream does not interact closely with the potential peak. Observational distinction in this regime is difficult unless the stream's impact parameter range covers the critical region near the peak, where the naked singularity produces a much broader scattering fan.
    \item We identified a transition at $q \approx 1.06$ where the centrifugal barrier disappears for the considered angular momentum. Above this threshold, the scattering is entirely dominated by the repulsive core, resulting in a monotonic deflection profile and narrow deflection angles for narrow infalling streams. This can mimic the reflection from a black hole for streams that do not interact with the potential barrier peak. However, a wide stream reveals a difference: the black hole captures a portion of the stream interacting with the peak, whereas the naked singularity scatters the entire stream---including the deeply penetrating material---potentially resulting in a wider angular distribution.
\end{enumerate}
}

In summary, the definitive dynamical signature of a Reissner-Nordström-like naked singularity is the total reflection of deeply plunging tidal streams that would otherwise be captured. Observationally, this would manifest as the presence of a returning debris stream where none is expected. A secondary, corroborating signature is the observation of a broad, widely-distributed fan of scattered material, pointing to an interaction with a centrifugal peak setting qualitative constraints in the charge-to-mass ratio (or angular momentum).

\section*{Acknowledgements}

This work was supported in part by the Polish NCN grant 2019/35/O/ST9/03965. MW is supported by a Ramón y Cajal grant RYC2023-042988-I from the Spanish Ministry of Science and Innovation {and acknowledges financial support from the Severo Ochoa grant CEX2021-001131-S funded by MCIN/AEI/ 10.13039/501100011033. AK acknowledges support from the NCN/NAWA PRELUDIUM BIS 1 internship (no. PPN/STA/2021/1/00099) to the Institute of Physics at the Silesian University in Opava in autumn 2024. AK thanks the hosting institute for their hospitality.}

\section*{Data Availability}

{The code used for the numerical computations and the generation of figures presented in this article is publicly available at \url{https://gitlab.camk.edu.pl/karakonang/geodesics}.}
The data used in the work presented in this article are available upon request to the corresponding author.

\bibliographystyle{mnras}
\bibliography{RNorbits}

\appendix

\section{Scattering from the repulsive core}\label{ap:abovePeak}

\begin{figure}
\centering
\begin{subfigure}{\columnwidth}
\includegraphics[width=\textwidth]{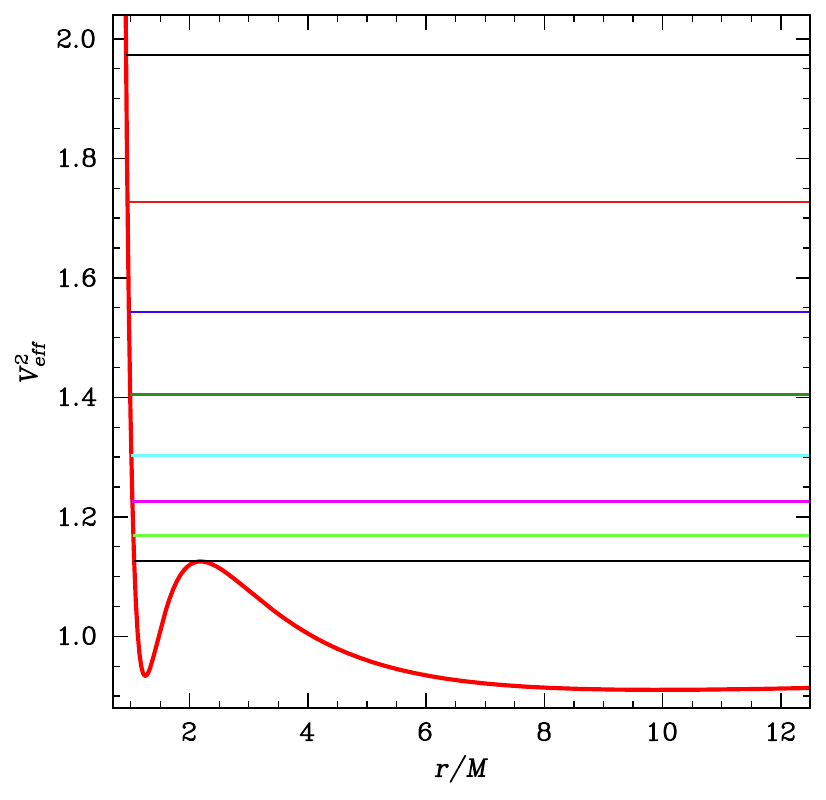}
\end{subfigure}
\hfil
\begin{subfigure}{\columnwidth}
\includegraphics[width=\columnwidth]{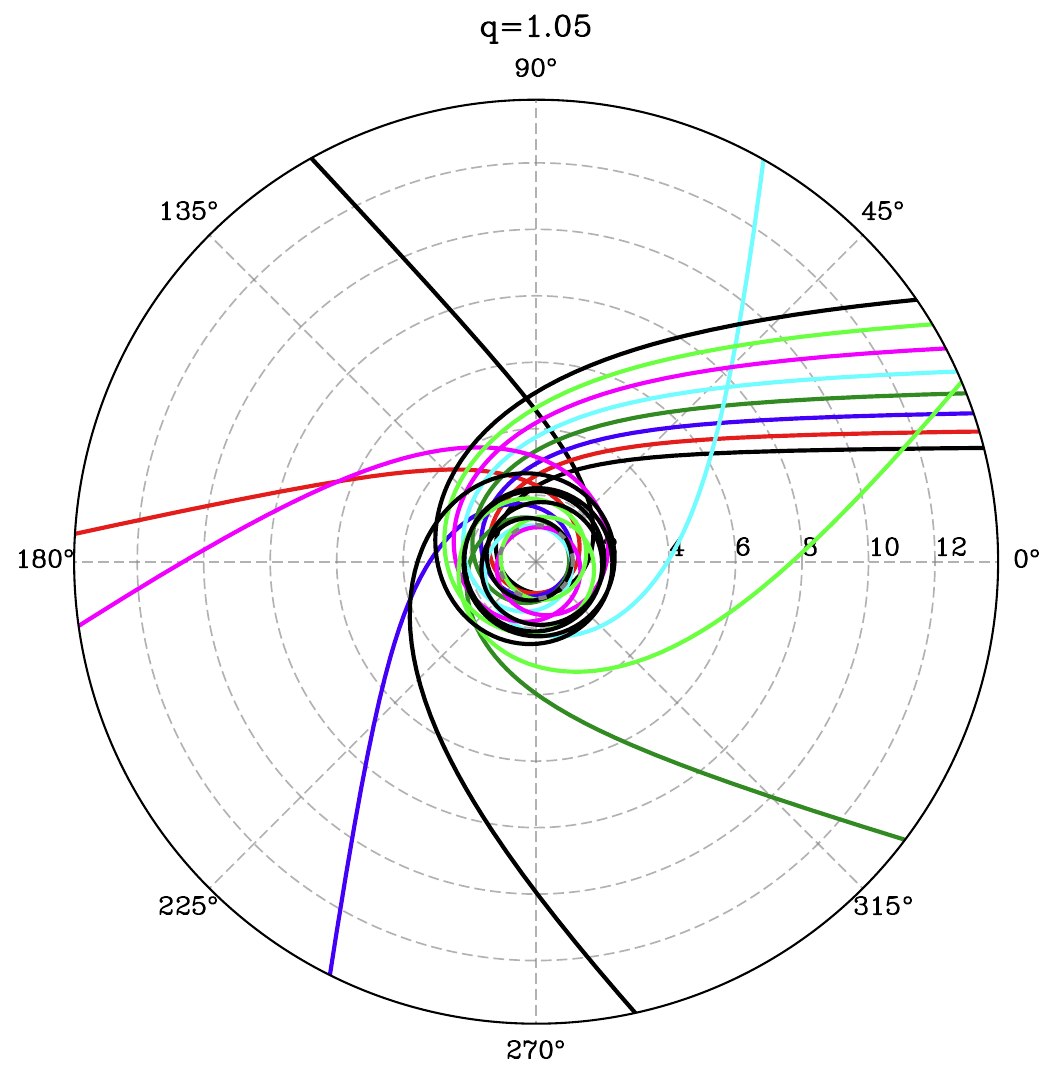}
\end{subfigure}
\caption{Trajectories around a NkS (\(q=1.05\)) for orbits initialized at low values of \(y_0\), corresponding to large values of \(\varepsilon\). The top panel presents the effective potential with horizontal lines indicating the value \(\varepsilon^2\) of orbits plotted in the bottom panel. The specific angular momentum is fixed to \(\ell=3.5M\).}
\label{fig:nearPeak}
\end{figure}

In this section, we examine the trajectories of particles with energies exceeding the peak of the centrifugal barrier around a naked singularity ($q=1.05$). Unlike the black hole case, where such high-energy particles are invariably captured, the naked singularity's repulsive core ensures their eventual reflection. Fig.~\ref{fig:nearPeak} illustrates this behavior for a range of energy values. The black trajectory represents a particle with energy just above the critical value $\varepsilon_{\rm crit}^2 = V_\ell^2(r_{\rm uco})$. This particle whirls around the central object multiple times as it approaches the unstable circular orbit at the peak of the potential, before finally crossing the barrier and being reflected by the repulsive core in the strong-gravity region near the singularity. As the energy increases further (higher horizontal lines in the top panel), the number of rotations decreases, and the particles are reflected more directly by the core, resulting in varying deflection angles that can span the entire $2\pi$ range.

\section{Different specific angular momentum values}\label{ss:lOrbits}

\begin{figure}
\centering
\includegraphics[width=0.85\columnwidth]{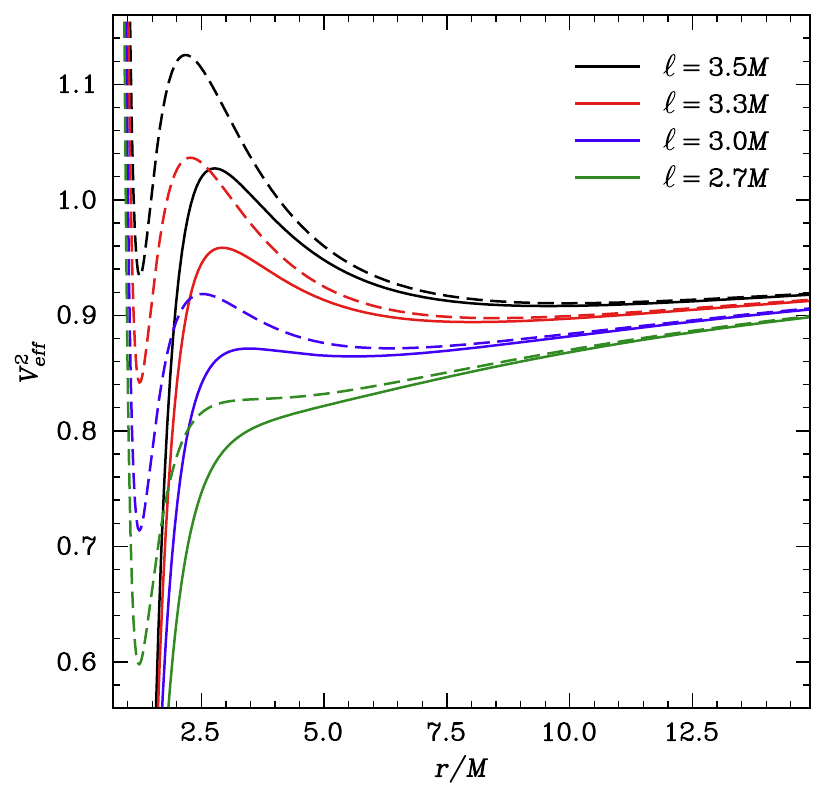}
\includegraphics[width=0.85\columnwidth]{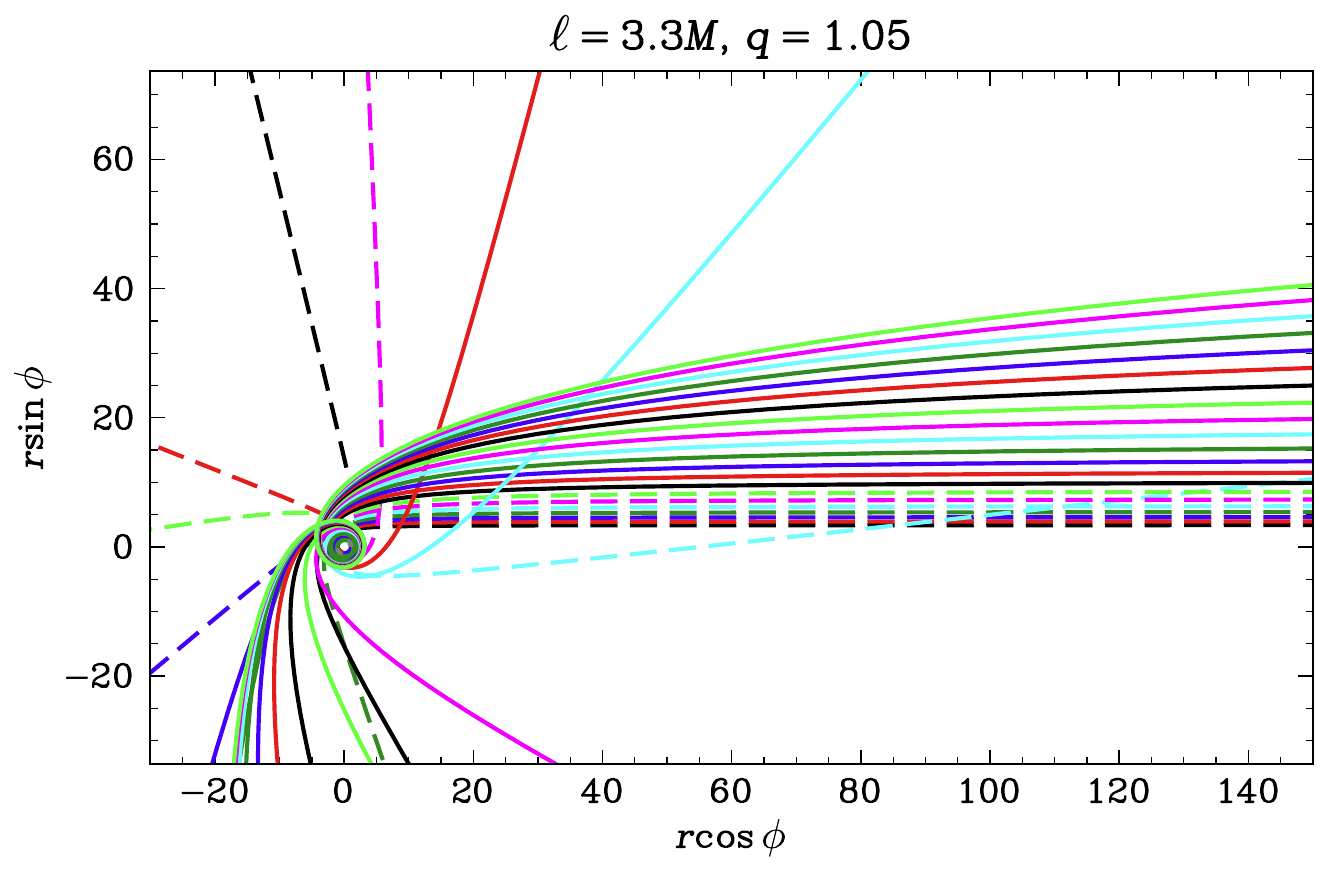}
\includegraphics[width=0.85\columnwidth]{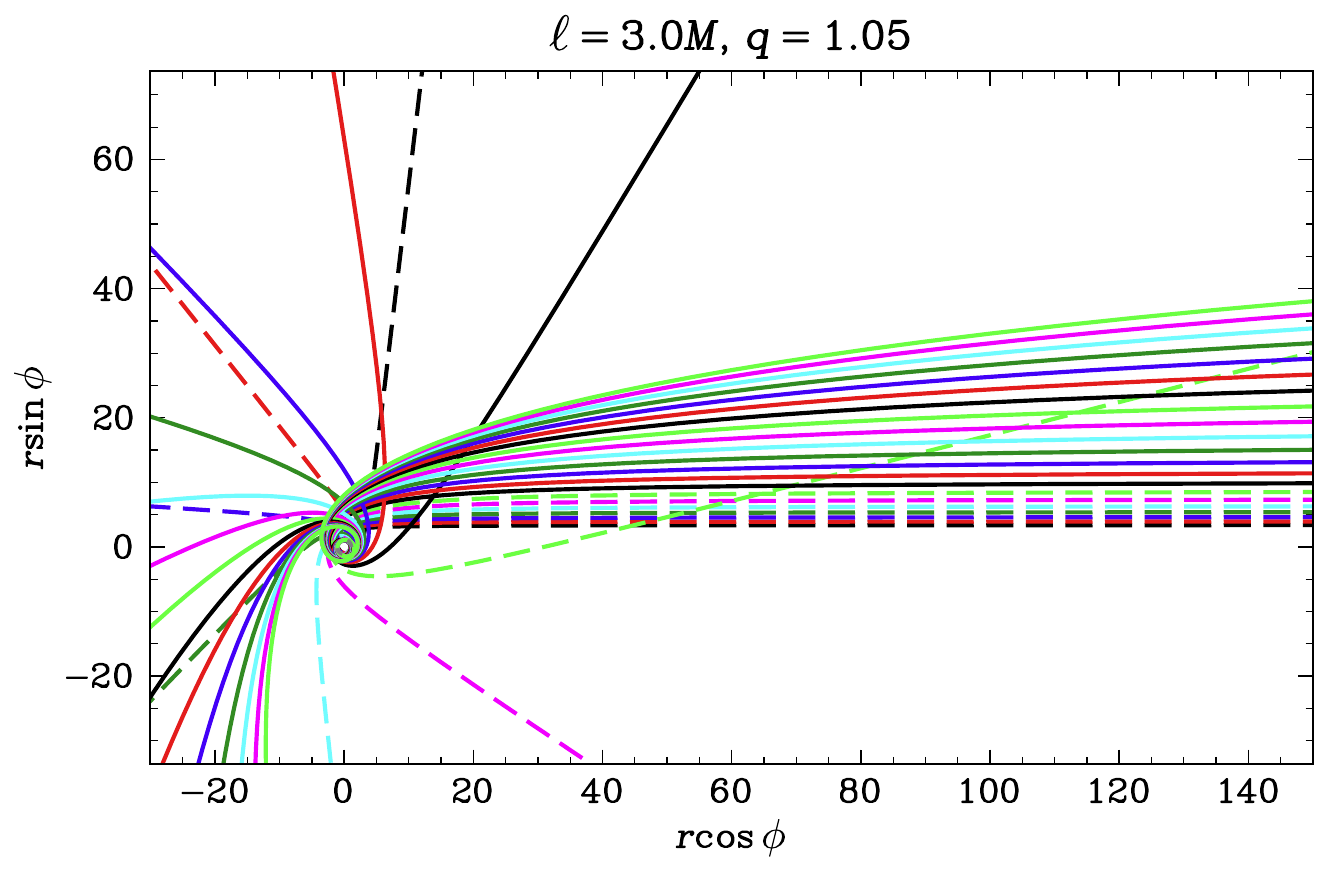}
\includegraphics[width=0.85\columnwidth]{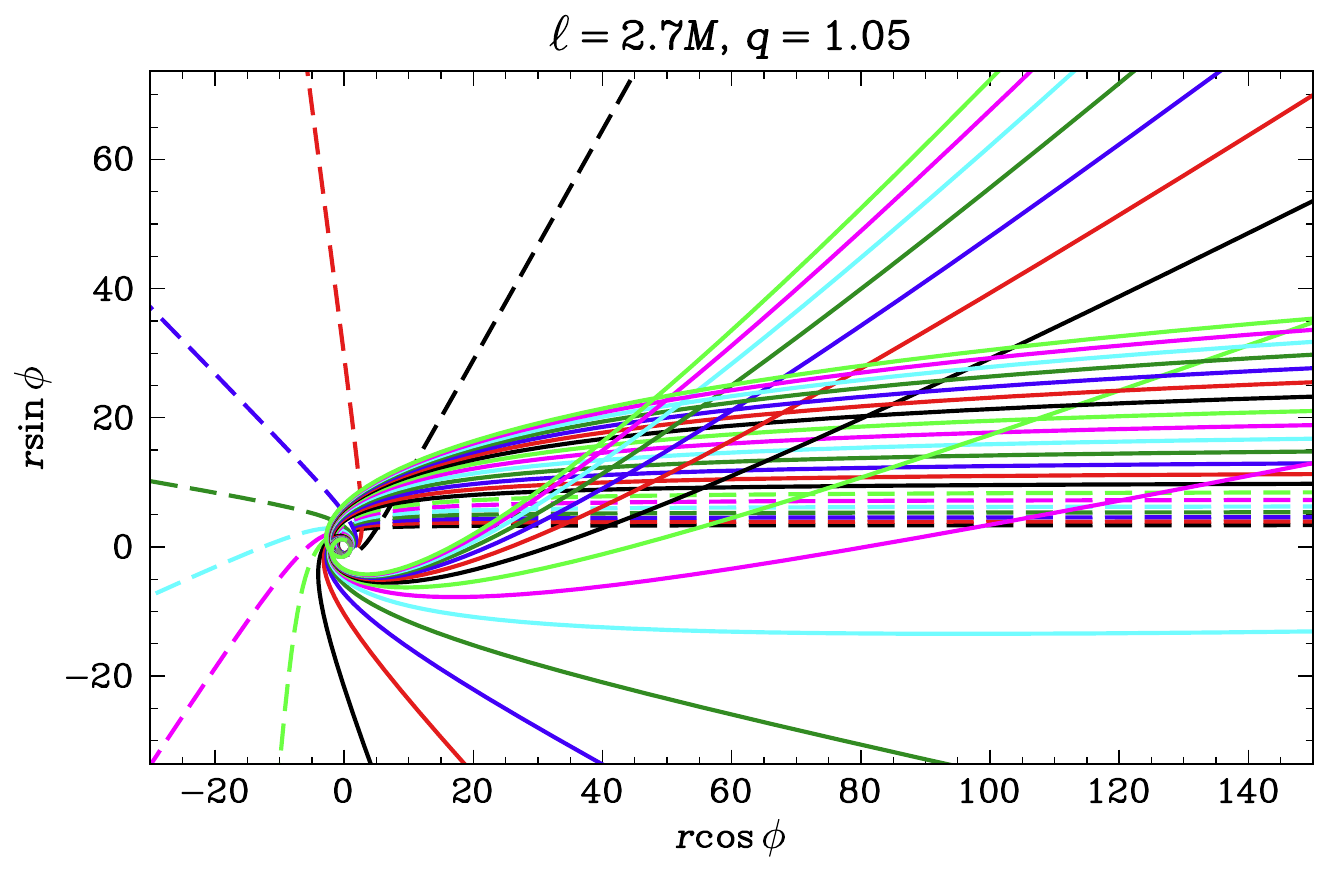}
\caption{The effect of specific angular momentum $\ell$ on the effective potential and trajectories around a naked singularity ($q=1.05$). Top panel shows the effective potential for various $\ell$ (solid lines: BH $q=0.95$, dashed lines: NkS $q=1.05$). The following three panels show the corresponding trajectories for $\ell/M = 3.3, 3.0$, and $2.7$, respectively. Dashed lines show low values in the initial impact parameter.}
\label{fig:appendix_combined}
\end{figure}

{
In this section, we repeat our previous investigations for additional values of the specific angular momentum \(\ell\). In the top panel of Fig.~\ref{fig:appendix_combined} we indicate the effective potential curves for the two compared spacetimes of a BH and of a NkS, with four distinct values of the specific angular momentum $\ell$ shown in different colors. For a low $\ell$, the maximum of \(V_\text{eff}^2\) corresponds to a value below 1. For a BH case this implies that all particles are absorbed, irrespectively of their $\varepsilon$. TDE debris in such a range of parameters would entirely disappear inside the BH, but it would be reflected and scattered by a NkS.
}

Bottom panels of Fig.~\ref{fig:appendix_combined} present trajectories around the NkS for particles with \( \ell = 3.3M, 3M, 2.7M\). The main difference between each panel and the results obtained with \(\ell=3.5M\) (Fig.~\ref{fig:cartOrbits}) is the angle of deflection and the radius of the closest approach. As shown in Fig.~\ref{fig:appendix_combined} (top panel), the peak of the potential barrier \(V_\text{eff}^2(r_{\rm uco})\) in NkS with $q = 1.05$ is below 1 for sufficiently small $\ell$. Thus, for the two lowest values of \(\ell\) considered (bottom rows of Fig.~\ref{fig:appendix_combined}), all particles arriving from infinity are reflected by the repulsive core around the NkS rather than by the centrifugal barrier. In Fig.~\ref{fig:appendix_combined} for $\ell = 3.3 M$, some particles are reflected by the centrifugal barrier and some by the repulsive NkS core related to the zero-gravity sphere.

\begin{figure}
    \centering
    \includegraphics[width=\columnwidth]{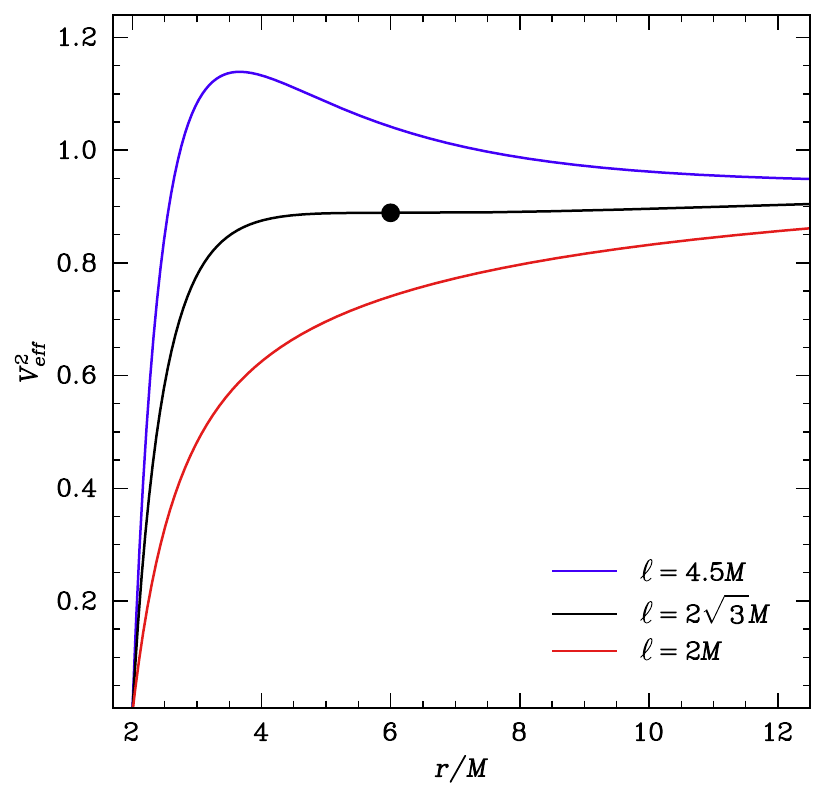}
    \caption{The Schwarzschild effective potential for particles with different specific angular momentum values. Curves correspond to decreasing values of specific angular momentum \(\ell\) from top to bottom. The limiting value in circular orbits of \(\ell=2\sqrt{3}M\) is shown with the black curve together with the location of ISCO (black dot).
    }
    \label{fig:vEff-Sch}
\end{figure}

\subsection{Parameter Study: Schwarzschild Black Hole}
\begin{figure*}
    \centering
    \begin{subfigure}{0.48\textwidth}
        \centering
        \includegraphics[width=\textwidth]{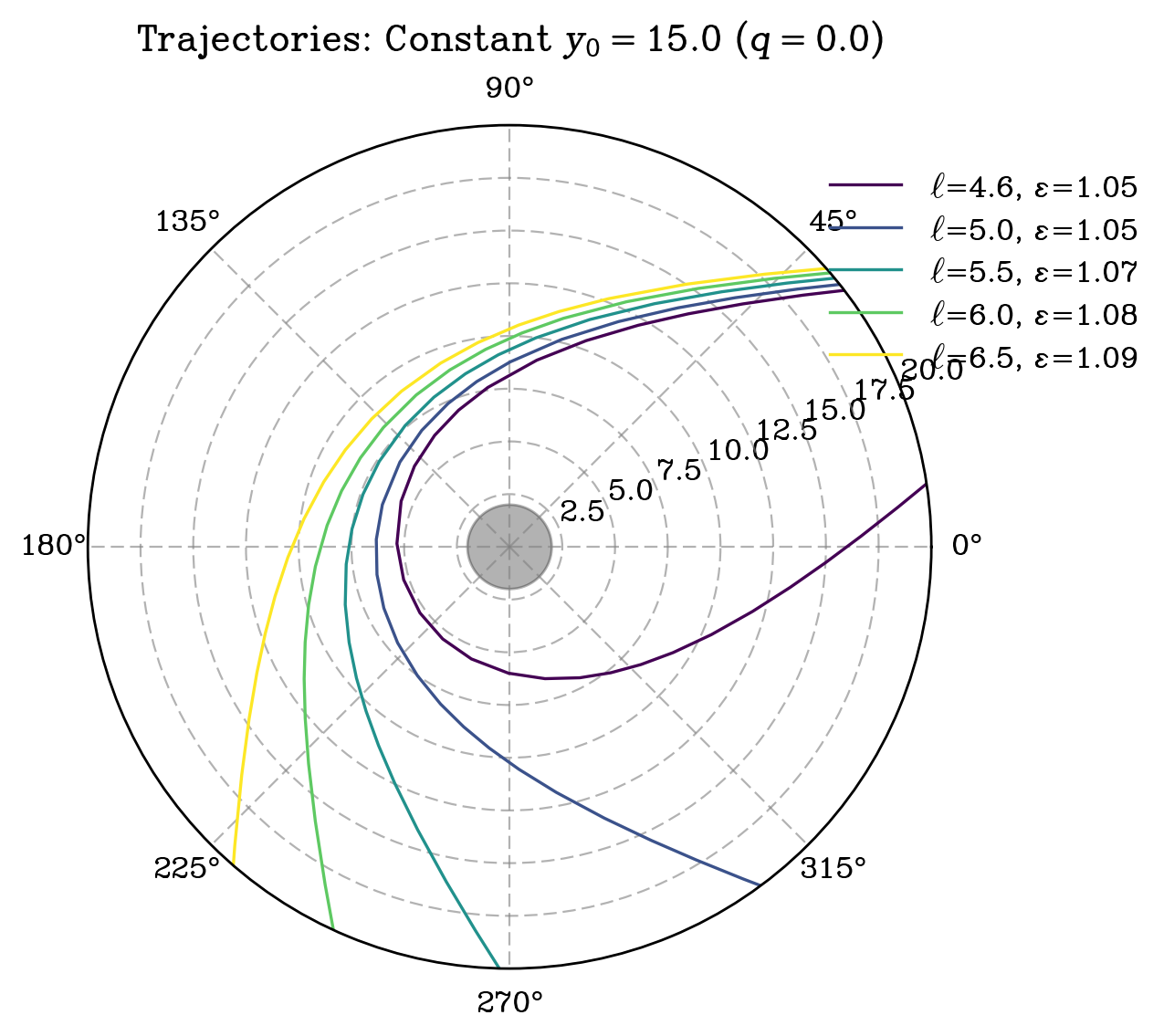}
        \caption{Trajectories ($y_0 = 15.0 M$)}
    \end{subfigure}
    \hfill
    \begin{subfigure}{0.48\textwidth}
        \centering
        \includegraphics[width=\textwidth]{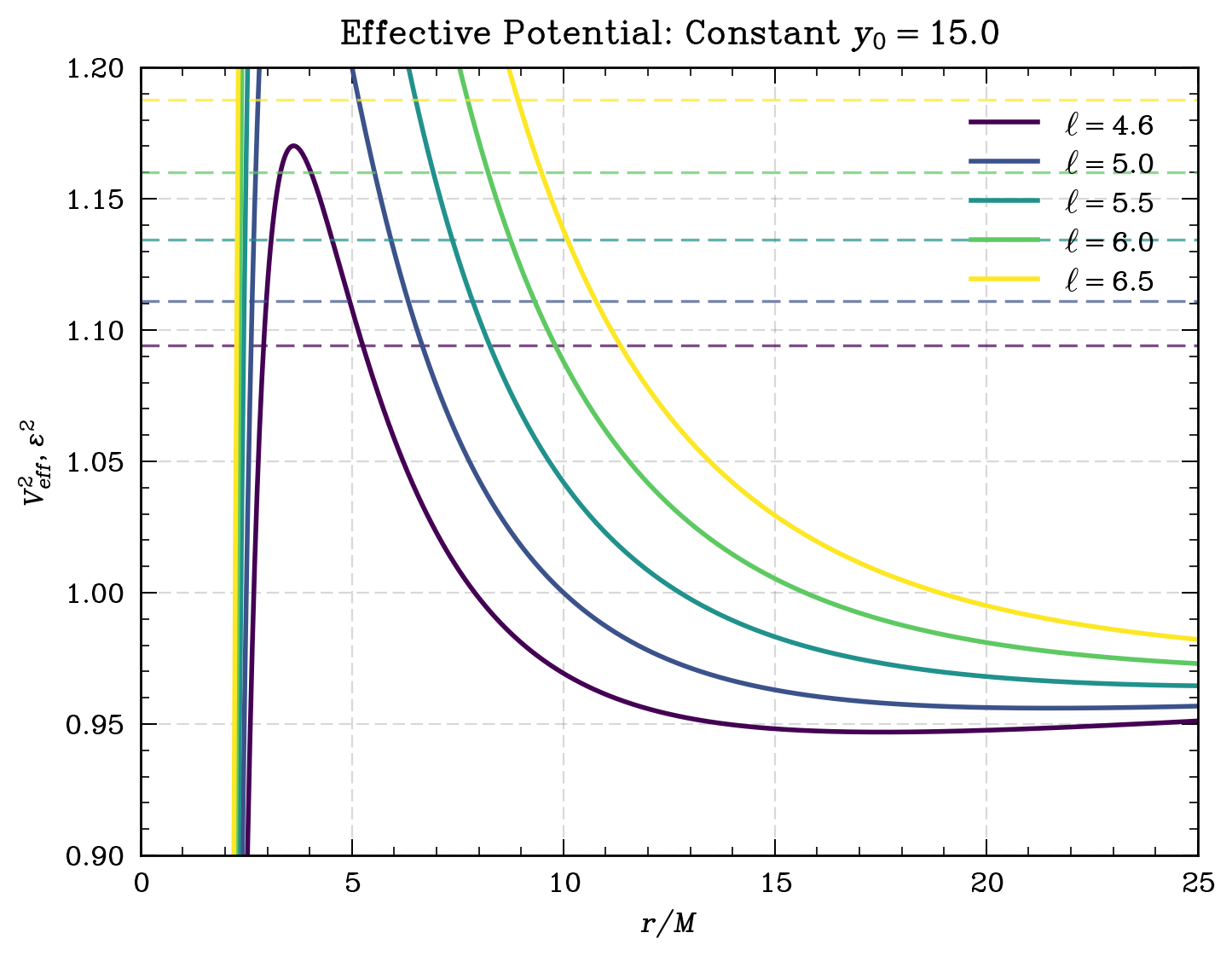}
        \caption{Effective Potential ($y_0 = 15.0 M$)}
    \end{subfigure}
    \medskip
    \begin{subfigure}{0.48\textwidth}
        \centering
        \includegraphics[width=\textwidth]{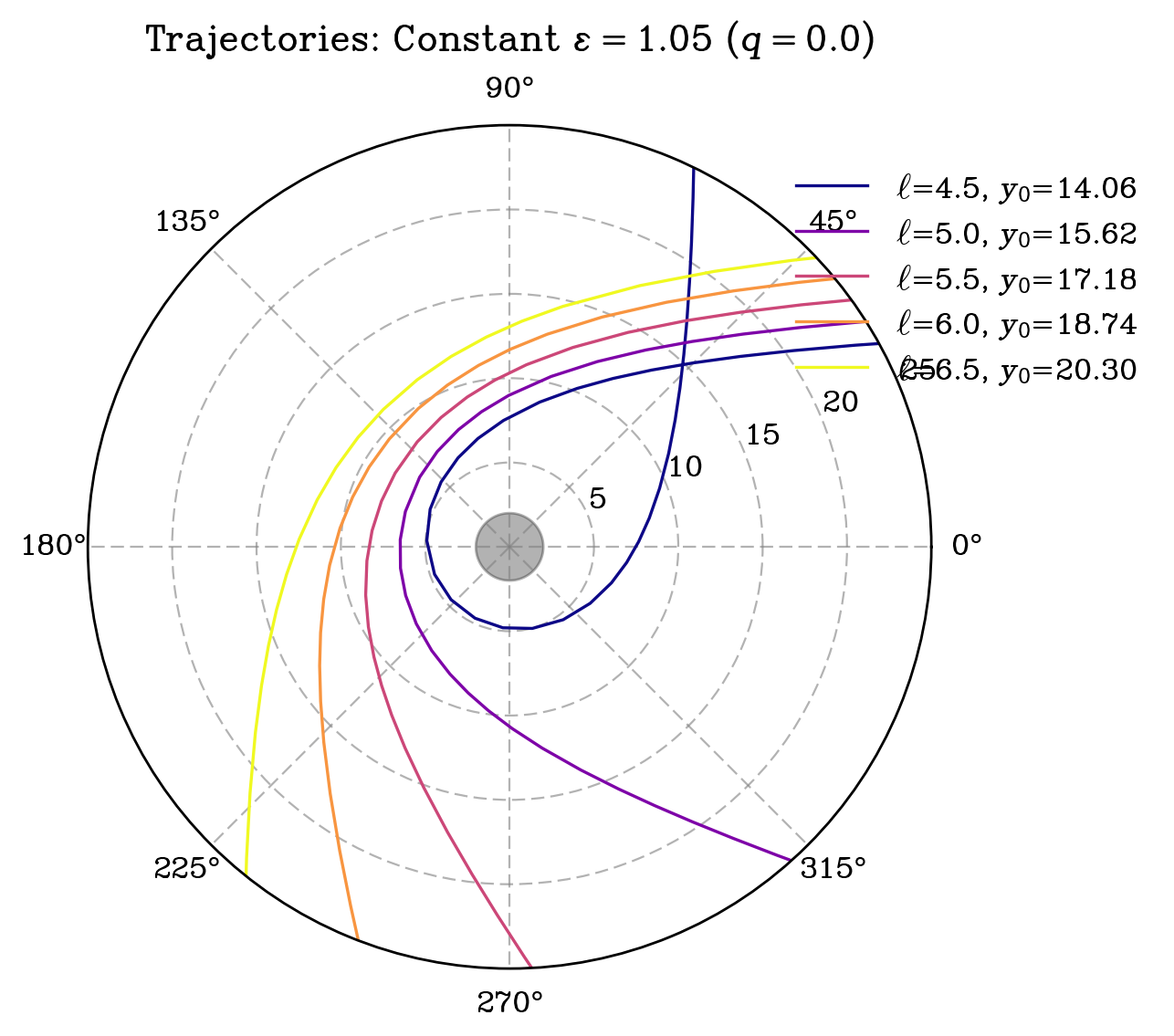}
        \caption{Trajectories ($\varepsilon = 1.05$)}
    \end{subfigure}
    \hfill
    \begin{subfigure}{0.48\textwidth}
        \centering
        \includegraphics[width=\textwidth]{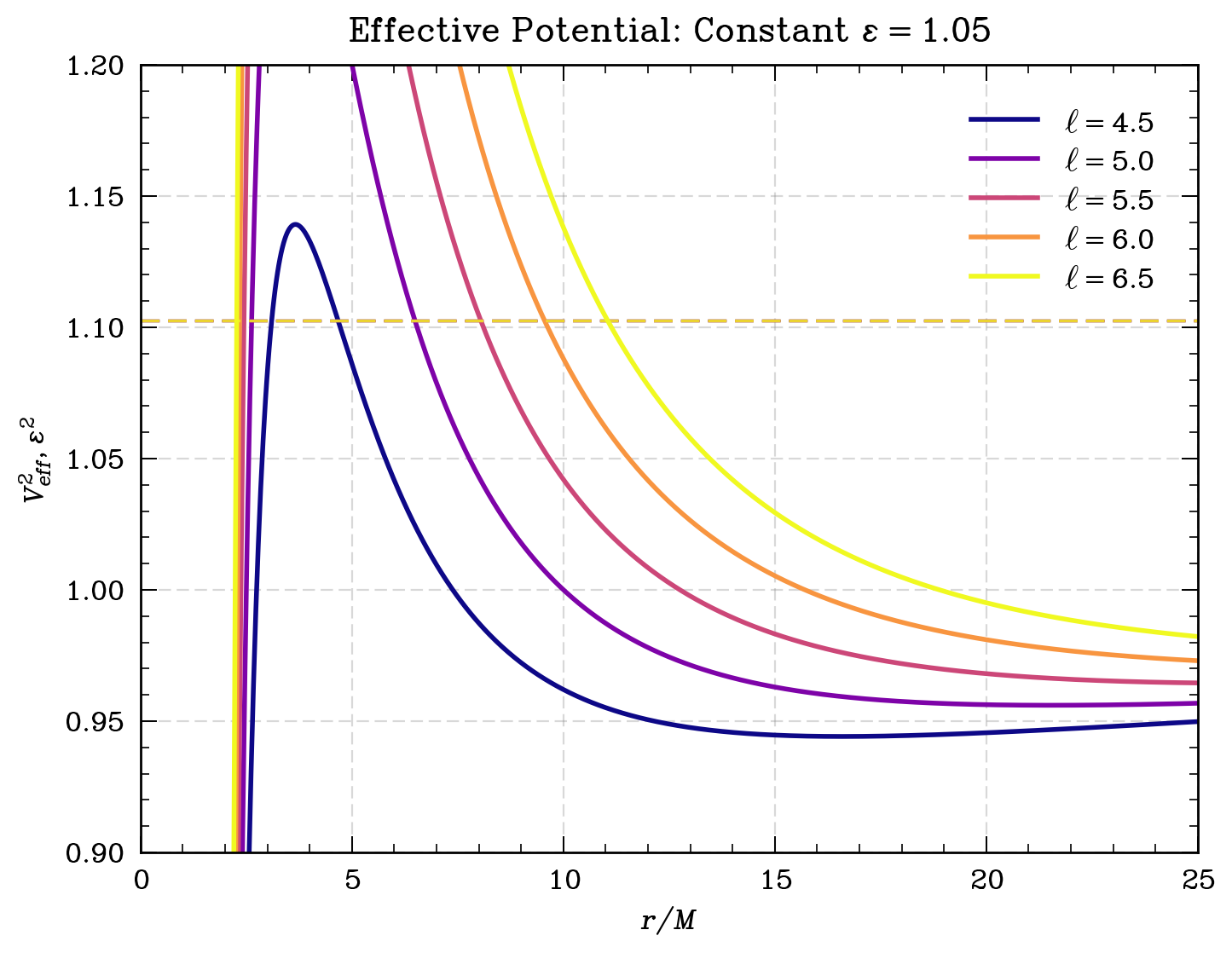}
        \caption{Effective Potential ($\varepsilon = 1.05$)}
    \end{subfigure}
    \caption{{Parameter study of trajectories and effective potentials in Schwarzschild spacetime ($q=0$) for $\ell \in [4.6, 6.5]$. \textit{(top row:)} Constant impact parameter $y_0 = 15.0 M$ (varying $\varepsilon$). \textit{(bottom row:)} Constant energy $\varepsilon = 1.05$ (varying $y_0$). Dashed lines in potential plots indicate the energy level corresponding to each angular momentum.}}
    \label{fig:schw_param_study}
\end{figure*}

{
We extended our analysis to the Schwarzschild black hole case ($q=0$) to investigate the sensitivity of scattering to variations in the specific angular momentum $\ell$ for high values ($\ell \ge 4.6 M$). Fig.~\ref{fig:schw_param_study} presents the trajectories for two scenarios.
}

{
In the first case (top panel), we keep the impact parameter fixed at $y_0 = 15.0 M$ and vary $\ell$. Since $y_0 = \ell / \sqrt{\varepsilon^2 - 1}$, varying $\ell$ at fixed $y_0$ implies changing the particle's energy. We observe that lower angular momenta (and thus lower energies) lead to deeper penetration and stronger scattering before the particle escapes.
In the second case (bottom panel), we fix the specific energy at $\varepsilon = 1.05$ and vary $\ell$. This corresponds to varying the impact parameter $y_0$. Here, we see the transition from scattered to (nearly) captured as $\ell$ decreases (and $y_0$ decreases).
}

\bsp	%
\label{lastpage}
\end{document}